\documentclass[10pt,journal,compsoc]{IEEEtran}
\usepackage{graphicx}
\usepackage{cite}
\usepackage{amsmath,amssymb,amsfonts}
\usepackage{algorithmic}
\usepackage{graphicx}
\usepackage{textcomp}
\usepackage[dvipsnames]{xcolor}
\usepackage{balance}
\usepackage[utf8]{inputenc}
\usepackage[T1]{fontenc}
\usepackage{url}
\usepackage{wrapfig}
\usepackage{caption}
\usepackage{algorithm}
\usepackage{hhline}
\usepackage{balance}
\usepackage{booktabs}
\usepackage{multicol}
\usepackage{multirow}
\usepackage{tabularx}
\usepackage{longtable}
\usepackage{array}
\usepackage{threeparttable}
\usepackage{wasysym}
\usepackage{colortbl}

\newcommand{\sectopic}[1]{\vspace{0em}\par\noindent{\textit{\bfseries #1}}}

\ifCLASSINFOpdf
\else
\fi
\begin{document}
%

\title{
AI-enabled Automation 
for  
Completeness Checking of Privacy Policies 
}


\author{Orlando~Amaral,~
        Sallam~Abualhaija,~\IEEEmembership{Member,~IEEE,}
        Damiano~Torre,~\IEEEmembership{Member,~IEEE,}
        Mehrdad~Sabetzadeh,~\IEEEmembership{Member,~IEEE,}
        and Lionel~C.~Briand,~\IEEEmembership{Fellow,~IEEE,}\\
\IEEEcompsocitemizethanks{
\IEEEcompsocthanksitem This work has been submitted to the IEEE for possible publication. Copyright may be transferred without notice, after which this version may no longer be accessible.
\IEEEcompsocthanksitem O.Amaral, S. Abualhaija, D. Torre, M. Sabetzadeh and L.C. Briand are with the SnT Centre for Security, Reliability, and Trust, University of Luxembourg, Luxembourg.\protect\\
E-mail: \{orlando.amaralcejas, sallam.abualhaija, mehrdad.sabetzadeh, lionel.briand\}@uni.lu
\IEEEcompsocthanksitem M. Sabetzadeh and L.C. Briand are also affiliated with the school of Electrical Engineering and Computer Science, University of Ottawa, Canada.\protect\\
E-mail: \{m.sabetzadeh, lbriand\}@uottawa.ca
\IEEEcompsocthanksitem D. Torre is also affiliated with the Department of Computer Information Systems, Texas A\&M University -- Central Texas, United States.\protect\\
E-mail: damiano.torre@tamuct.edu
}
\thanks{Manuscript received Month DD, 2021; revised Month DD, 2021.}}


%


\markboth{Journal of \LaTeX\ Class Files,~Vol.~14, No.~8, August~2015}%
{Shell \MakeLowercase{\textit{et al.}}: Bare Demo of IEEEtran.cls for Computer Society Journals}

\IEEEtitleabstractindextext{%
\begin{abstract}

Technological advances in information sharing have raised concerns about data protection. 
Privacy policies contain privacy-related requirements about how the personal data of individuals will be handled by an organization or a software system (e.g., a web service or an app). 
In Europe, privacy policies are subject to compliance with the General Data Protection Regulation (GDPR). 
A prerequisite for GDPR compliance checking is to verify whether the content of a privacy policy is complete according to the provisions of GDPR. Incomplete privacy policies might result in large fines on violating organization as well as incomplete privacy-related software specifications. 
Manual completeness checking is both time-consuming and error-prone. 
In this paper, we propose AI-based automation for the completeness checking of privacy policies. Through systematic qualitative methods, we first build two artifacts to characterize the privacy-related provisions of GDPR, namely a conceptual model and a set of completeness criteria. Then, we develop an automated solution on top of these artifacts by leveraging a combination of natural language processing and supervised machine learning. Specifically, we identify the GDPR-relevant information content in privacy policies and subsequently check them against the completeness criteria. 
To evaluate our approach, we collected 234 real privacy policies from the fund industry. Over a set of 48 unseen privacy policies, our approach detected 300 of the total of 334 violations of some completeness criteria correctly, while producing 23 false positives. The approach thus has a precision of 92.9\% and recall of 89.8\%. Compared to a baseline that applies keyword search only, our approach results in an improvement of 24.5\% in precision and 38\% in recall.

\end{abstract}

}

\maketitle
\IEEEdisplaynontitleabstractindextext

%
\IEEEpeerreviewmaketitle

\section{Introduction} \label{sec:introduction}

\textcolor{black}{Advances in information sharing technologies have raised concerns about protecting the privacy of individuals. }
In Europe and indeed worldwide, the General Data Protection Regulation (GDPR)~\cite{EU2018} is widely viewed as a benchmark for data protection and privacy regulations. GDPR harmonizes data privacy laws across the European Economic Area (EEA), providing further protection to individuals for controlling their personal data in the face of new technological developments~\cite{EUP2019}. 

While undoubtedly beneficial to individuals in many ways, the reality is that organizations are having considerable difficulty complying with GDPR~\cite{Tankard2016}. 
There is thus a pressing need for cost-effective methods that can help different organizations better deal with privacy considerations. This need has not gone unnoticed by the research community. For example, Perrera et al.~\cite{Perera2020} propose systematic guidance to help software engineers develop privacy-aware applications; Torre et al.~\cite{Torre19,Torre2020GDPRModel} propose the use of Model-driven Engineering as a basis for GDPR compliance automation; and Ayala-Rivera and Pasquale~\cite{Ayala2018} present a step-wise approach for eliciting requirements related to GDPR compliance.

To comply with GDPR, organizations need to take into account the principles of personal data processing set out in the regulation, and to regularly review their measures, practices and processes related to the collection, use and protection of personal data. 
\textcolor{black}{Compliance also entails that software systems storing or processing personal data should properly implement privacy-related GDPR requirements.}
Every organization, whether \textcolor{black}{Europe}-based or not, which is collecting, processing or in some way handling the personal data of \textcolor{black}{European} citizens and residents \hbox{must comply with GDPR.} 

In this paper, we concern ourselves with GDPR \emph{privacy policies}.
A privacy policy can be viewed as a technical document stating the multiple privacy-related requirements  that an organization (including processes, services, developed systems) should satisfy in order to help users make informed decisions about the data that this organization may collect and use. In other words, a privacy policy explains how an organization handles personal data and how it applies the principles of GDPR. Privacy policies are usually defined  through natural-language statements. Natural language (NL) is an ideal medium for expressing privacy policies since it is flexible and universal~\cite{Caramujo2019}. 
\textcolor{black}{Though NL is advantageous for establishing a common understanding,  processing NL documents is challenging due to common quality issues such as ambiguity, incompleteness and inconsistency~\cite{BhatiaB18}. }

This paper tackles an important dimension of GDPR compliance checking for privacy policies. Specifically, in collaboration with legal experts from Linklaters (a multinational law firm), we develop an AI-enabled approach for checking whether a given privacy policy is ``complete'' according to the provisions of GDPR. 
We use the term ``complete'' rather than ``compliant'' to signify the fact that our approach can detect only the presence (or absence) of the information content types that GDPR envisages for privacy policies. 
\textcolor{black}{A privacy policy is deemed ``complete'' according to GDPR if it explicitly contains certain information mandatory for ensuring data protection and privacy rights, e.g., about the rights individuals have over their personal data. We further clarify this concept with an example in the next section.}
\textcolor{black}{
According to our collaborating legal experts, \emph{completeness checking} is an essential prerequisite for compliance checking. Manually checking completeness is however both time-consuming and error-prone. Providing automated support is thus desirable so that legal experts can focus their effort on more critical tasks. }

\subsection{Practical Scenario}

In practice, completeness checking of privacy policies against GDPR can be beneficial to a diverse group of legal experts, software engineers, and other business stakeholders. 
The first step in completeness checking is to determine if GDPR-relevant information content is present or not in a given privacy policy. 
Based on the above analysis, the second step is then to map what is actually present in the privacy policy to what must be present according to the provisions of GDPR.
In the rest of this paper, we will use the term \textcolor{black}{\textit{metadata type} to describe any information type which we extracted from the privacy-related provisions of GDPR. Some of these metadata types are mandatory and thus have a direct impact on completeness. 
We elaborate how we combine the metadata types for checking the completeness of a privacy policy in Sec.~\ref{sec:criteria}.} 
A comprehensive description of these metadata types is provided in Sec.~\ref{sec:qualitative}. 

\begin{figure*}[!t]
\centering
\includegraphics[width=\linewidth] {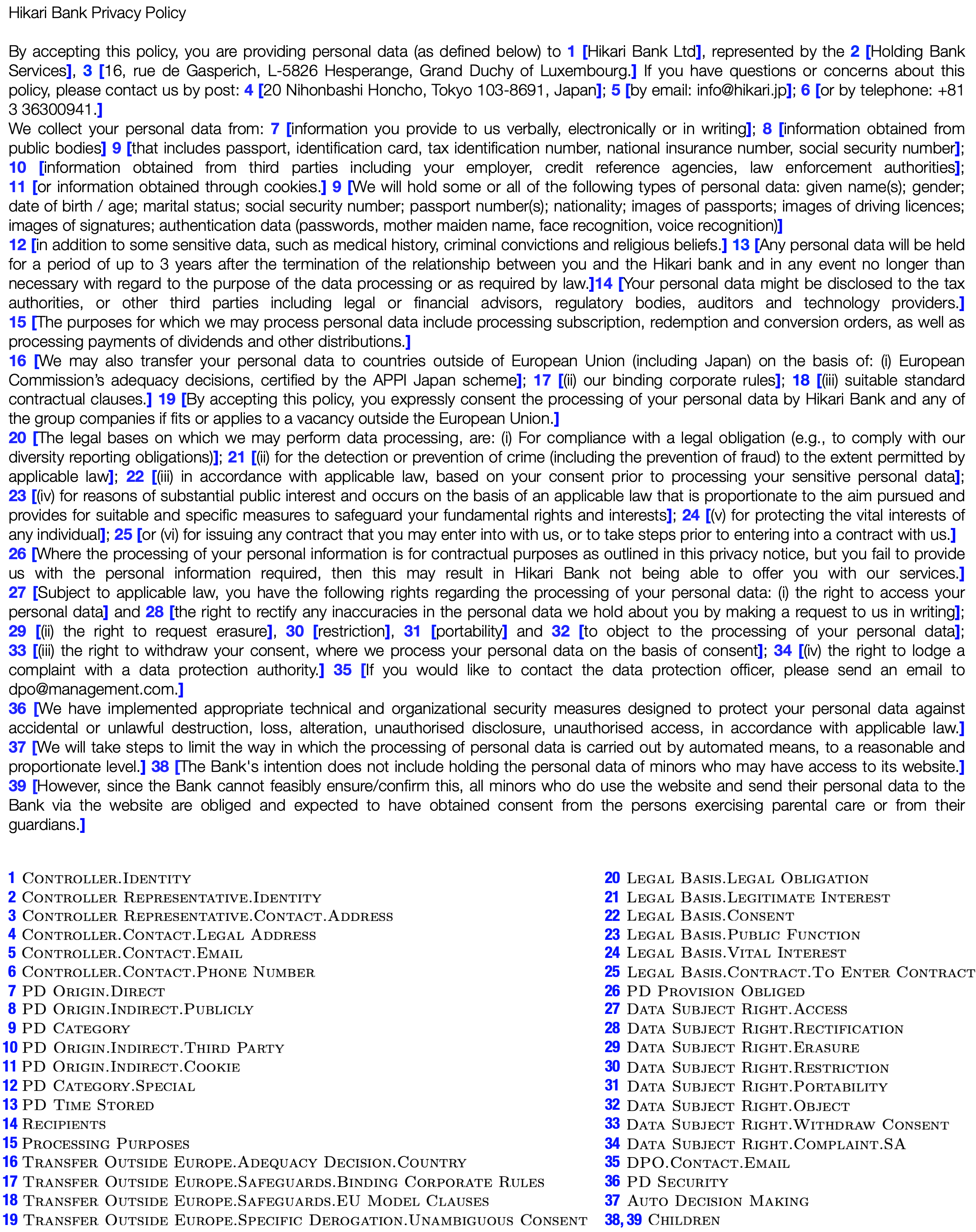}
\caption{Example of a fully annotated privacy policy.} \label{fig:annotations}
\end{figure*}

Examples of metadata types include: \textsc{Processing Purposes} to characterize the purposes of the processing for which personal data is being collected, 
\textsc{Legal Basis} to capture the legal basis for the processing of personal data, \textcolor{black}{and \textsc{Data Subject Right} to mark the clause(s) giving an individual the rights in relation with their personal data. Under \textsc{Data Subject Right}, several specializations are listed to describe the different rights an individual has. For instance, \textsc{Data Subject Right.Access} is concerned with the right to request access to the personal data from the controller. The specializations of the metadata types are represented, throughout the paper, with a \textit{dot}. 
Fig.~\ref{fig:annotations} shows a \textcolor{black}{complete privacy policy that is annotated with all metadata types (from Sec.~\ref{sec:qualitative}).} In the figure, we present the metadata types using numbers (further explained in the legend), and square brackets to delineate the text corresponding to the metadata types.} 
For example, number \textbf{15} in Fig.~\ref{fig:annotations} refers to the metadata type \textsc{Processing Purposes}, numbers \textbf{20 -- 25} refer to different specializations of \textsc{Legal Basis}, and number \textbf{27} refers to \textsc{Data Subject Right.Access}.

\textcolor{black}{To deem the example privacy policy in Fig.~\ref{fig:annotations}  complete, GDPR requires the presence of multiple mandatory metadata types, including the ones concerning \textsc{Controller}, i.e., the organization which collects personal data (GDPR, Art. 13 and Art. 14(f)). In particular, the policy should include the identity (i.e., \textsc{Controller.Identity}) and contact details of the controller (i.e., \textsc{Controller.Contact}). }
As we see in Fig.~\ref{fig:annotations}, these two metadata types are  mentioned respectively in number~\textbf{1}, 
and numbers \textbf{4} -- \textbf{6}. 
\textcolor{black}{Verifying the presence of the metadata types about \textsc{Controller} is however not sufficient, and  verification of other metadata types is needed in order to make the final decision as to whether the privacy policy is complete according to GDPR. }  

\textcolor{black}{Legal provisions in GDPR can contain requirements which depend on one another. Consequently, the presence of certain metadata types in a privacy policy may necessitate the presence of certain other metadata types in the policy. } 
For instance, if a privacy policy states that the legal basis for the processing of personal data is based on individual consent (i.e.,  \textsc{Legal Basis.Consent}), then the right to withdraw this consent should be granted in the same policy (i.e., \textsc{Data Subject Right.Withdraw Consent}). These metadata types correspond to two different GDPR articles, Art. 6.1(a) and Art. 13.2(c), respectively. 
Information related to these types can be found by reviewing paragraphs that are usually located in different parts of the privacy policy. In Fig.~\ref{fig:annotations}, \textsc{Legal Basis.Consent} is mentioned in the text: \textit{in accordance with applicable law, based in your consent [...]} (number~\textbf{22}), while \textsc{Data Subject Right.Withdraw Consent} is mentioned in: \textit{the right to withdraw your consent [...]} (number~\textbf{33}). 
If done manually, this back-and-forth reviewing of the text requires a considerable amount of effort and time in practice.

Checking the completeness of a given privacy policy according to GDPR is essential for ensuring the completeness of the privacy-related software requirements induced by the policy. 
To illustrate, consider our example in Fig.~\ref{fig:annotations}. 
Since the \textsc{Controller} (i.e., Hikari Bank Ltd -- number \textbf{1}) is located in Japan, it is likely that the personal data of the bank's customers will be transferred outside the \textcolor{black}{Europe}. Articles 13.1(a), 13.1(f) and 14.1(f) in GDPR enforce requirements to ensure the protection of personal data, for example when transferred outside \textcolor{black}{Europe}.
The implications of these articles are then two-fold. On one hand, the privacy policy must provide information about the \textsc{Identity} and \textsc{Contact} details of the \textsc{Controller Representative} (who has to be located in Europe) -- as shown in numbers \textbf{2} and \textbf{3}, respectively. The policy must also state the legal agreement that is in place for transferring data to Japan such as the Act on the Protection of Personal Information (APPI) --  provided in number \textbf{16}. 
On the other hand, the software developed to handle such personal data (e.g., the online banking service) has to comply with Japan’s data protection law. 
APPI-compliant software should provide a response to the individuals' requests in relation with their personal data within two weeks. Otherwise, an individual can sue the controller. 
An incomplete privacy policy due to the missing metadata types related to the location where the controller is based (in our case, Japan) or to the legal agreement used for transferring data (i.e., APPI) fails to comply with GDPR. The missing metadata types will remain unknown for a software developer and might lead to developing a non-compliant system. Consequently, the organization could bear significant fines for violating data-protection rules.

More precisely, a privacy policy can be considered as a form of legally binding requirements specification which describes some of the properties and functionalities of a system-to-be. 
Therefore, completeness checking of privacy policies, and identifying their metadata as a primary step, can be seen as part of a broader solution to ensure legal compliance in information systems. In the software engineering (SE) literature, there have been attempts at mapping the text of a privacy policy to the implementation of a given software application, as a method for detecting GDPR violations~\cite{Slavin16, Fan20}. \textcolor{black}{For instance, based on what we argued earlier, the privacy-related requirement about answering an individual's request pertaining to their personal data has to be mapped onto some function in the developed software. }

Similarly, other metadata types identified in privacy policies can play a major role in software development. Examples include \textsc{PD security}, \textsc{PD Time stored}, \textsc{Data Subject Right.Erasure},  
\textsc{Legal Basis.Consent}, and \textsc{Data Subject Right.Withdraw Consent}. 
In response to \textsc{PD security}, the controller has to implement appropriate protection mechanisms during software development (e.g., using encryption) to avoid penalty charges for information leakage as stated in GDPR. 
Further, a software system has 
to automatically delete collected personal data according to the time limit specified in the privacy policy (\textsc{PD Time stored}) or upon an individual's request (\textsc{Data Subject Right.Erasure}). 
When the consent of an individual is required for processing personal data (\textsc{Legal Basis.Consent}), a software system has to implement a clear request procedure for consent where the individual takes an action to provide consent, e.g., by checking an ``I agree'' checkbox. As stipulated by GDPR, the system would also have to provide individuals with the possibility to withdraw this consent (\textsc{Data Subject Right.Withdraw Consent}). 
\textcolor{black}{
The above examples show the benefits of 
completeness checking in  different scenarios. Since checking completeness manually  is time-consuming and effort-intensive, computer-assisted support for this task is advantageous.}

A naive completeness-checking solution is to automatically find certain metadata types in a privacy policy through searching for keywords that are commonly used to express these metadata types. 
Relying merely on keyword search is problematic due to several reasons. First, there are overlapping keywords among multiple metadata types. For example, the keyword ``protect'' can indicate three metadata types related to security, data protection office, and safeguards for transferring personal data outside of Europe. 
Second, some metadata types cannot be captured via keywords. For instance, the metadata type \textsc{Recipients} (i.e., the parties with which individual personal data is shared) is usually expressed in the privacy policy as a list of diverse organizations (number \textbf{14} in Fig.~\ref{fig:annotations}). Since each privacy policy can have a different list of  \textsc{Recipients}, using keyword search is infeasible for identifying this metadata type. \textcolor{black}{To illustrate, let us suppose that ``third parties'' is used as a keyword for identifying \textsc{Recipients}. Note that the same keyword can also be used to identify \textsc{PD Origin.Indirect.Third-party}. Searching for this keyword will result in missing all occurrences of \textsc{Recipients} that do not contain the keyword and falsely identifying some occurrences due to overlapping keywords. }
In addition to the limitations of keyword search, the problem of checking completeness raises several other challenges. A particular sentence can discuss one or more metadata types which can be described in a hierarchy based on the specializations introduced in GDPR. In other words, an automated solution should be able to predict multiple (hierarchical) labels (metadata types) for a given sentence in the privacy policy.
Inter-dependent metadata types (e.g., \textsc{Consent} and \textsc{Withdraw Consent}  --  discussed earlier) do not always occur consecutively in the privacy policy. This means that successful completeness checking requires identifying all the related metadata types accurately.

\subsection{\textcolor{black}{Research Questions}}
The paper investigates the following six research questions (RQs):

\textbf{RQ1: What are the metadata types required for checking the completeness of a privacy policy according to GDPR?} We answer RQ1 by building a conceptual model that specifies GDPR's information requirements for privacy policies. Our conceptual model, comprised of 56 metadata types, was developed in close collaboration with subject-matter experts. The concepts in this model are described in a glossary and are further traceable to the articles of GDPR. 

\textcolor{black}{\textbf{RQ2: What are the criteria for checking whether a privacy policy is complete according to GDPR?}}
Drawing on our conceptual model, to answer RQ2, we define a set of 23 criteria 
specifying what in a privacy policy should be checked for completeness against GDPR. 
Violating any of these criteria might lead to an incomplete privacy policy. 

\textcolor{black}{\textbf{RQ3: How can  privacy policies be automatically checked for completeness against GDPR?}
To answer RQ3, we use a combination of NLP and ML methods based on word embeddings and semantic similarity to develop an AI-based approach. Our approach identifies the different metadata types (from our conceptual model in RQ1) that are present in a privacy policy (\textit{metadata identification}), and then checks these metadata against the completeness criteria  (derived in RQ2) using automated conditional expressions (\textit{completeness checking}).
}

\textcolor{black}{\textbf{RQ4: How accurate is 
our proposed approach in identifying GDPR-relevant metadata in privacy policies?}
RQ4 examines the accuracy of our metadata identification approach.  
As we discuss in Sec.~\ref{sec:evaluation}, we achieve an average precision of 92.1\% and average recall of 95.3\% on an evaluation set made up of 48 unseen privacy policies. 
}

\textcolor{black}{\textbf{RQ5: How accurate is 
our approach in checking the completeness of privacy policies?} In RQ5, we investigate the accuracy of our automated approach in checking the completeness of privacy policies according to the provisions of GDPR.
Over the evaluation set, 
our approach successfully finds 300 out of 334 violations of the completeness criteria, while raising false alarms (false positives) in 23 cases. Our approach has thus a precision of 92.9\% and a recall of 89.8\%.}

\textcolor{black}{\textbf{RQ6: Is our approach worthwhile compared to a simpler solution?} In RQ6, we compare our AI-based approach to a  baseline that uses only keyword search. Compared to this baseline and over our evaluation set,  
using AI-technologies improves the metadata identification by an average precision of 26.9\% and average recall of 5.2\%. Our approach significantly improves the overall completeness checking of privacy policies by an average precision of 24.5\% and average recall of 38\%.}

\textcolor{black}{The research presented in this paper is an extension of a previous conference paper~\cite{TorreRE2020} published at the 28th IEEE International Requirements Engineering conference (RE’20). 
The current paper provides a much more extensive empirical investigation in terms of the research questions, privacy policies used for evaluation, and metadata types covered by these policies. 
\textcolor{black}{In particular, 
(1) we provide, through a concrete and detailed example, different scenarios where automated completeness checking turned out to be useful to a diverse group of people including lawyers and software engineers;
} 
(2) we include two more research questions: RQ2 for addressing the qualitative methods 
leading to the derivation of completeness criteria and RQ6 for comparing our approach to a simple, intuitive baseline; (3) we apply our AI-based approach for identifying all the 56 metadata types in a given privacy policy; to put this into perspective, our earlier conference paper dealt with only 20 metadata types; and (4) we improve our validation method to empirically evaluate our approach 
on 48 unseen privacy policies ($\approx$20\% of the entire dataset), instead of only 24 policies as was the case in our earlier conference paper. 
}

\subsection{\textcolor{black}{Contributions}}
This paper makes the following four contributions: 
 
(1) We  develop a conceptual model to characterize the content of privacy policies, as stated in the provisions of GDPR. This conceptual model provides an abstract and yet precise set of metadata types that one can expect to find in privacy policies according to GDPR. 

(2) We create a set of completeness criteria that describe when a privacy policy is considered complete according to GDPR. 
\textcolor{black}{For creating these criteria (and also the conceptual model in (1)), we use systematic qualitative methods, as will be further explained in the paper.  }

(3) We develop an automated completeness checking tool using AI technologies. 
Specifically, we devise an approach based on Natural Language Processing (NLP) and Machine Learning (ML) for automatically identifying the content of a given privacy policy. To do so, we rely on the metadata types in the conceptual model developed in (1) as classification types. 
Given the identified metadata, we subsequently \textcolor{black}{use the completeness criteria created in (2)} to automatically check whether a given policy meets the information requirements envisaged by GDPR. 

(4) We empirically evaluate our approach using a dataset of 234 privacy policies. 
\textcolor{black}{These policies collectively contain 19847 
sentences manually assigned (when applicable) to one or more of the metadata types from our conceptual model. The large majority (87\%) of these assignments have been made by independent, third-party annotators
(non-authors). 
} 
We use $\approx$80\% of our dataset for developing our proposed solution 
and the remaining $\approx$20\% for evaluation. On our evaluation set, our AI-based approach yields an average precision of 92.1\% and  average recall of 95.3\% in automatically identifying the different metadata types. Our completeness checking yields an average precision of 92.9\% and an average recall of 89.8\%. Compared to a baseline that uses keyword search, our approach leads to an overall average
improvement of 24.5\% in precision and 38\% in recall when checking the completeness of privacy policies.   

\textcolor{black}{\subsection{Structure}}
Sec.~\ref{sec:background} provides background information. 
Sec.~\ref{sec:qualitative} presents the qualitative study we conducted for building our privacy-policy conceptual model. Sec.~\ref{sec:criteria} describes the methods we used to create a set of  criteria for checking the completeness of privacy policies according to GDPR. Sec.~\ref{sec:approach} explains our proposed AI-based approach for checking the completeness of a given privacy policy. Sec.~\ref{sec:evaluation} discusses the empirical evaluation of our 
approach.  Sec.~\ref{sec:threats} discusses threats to validity. Sec.~\ref{sec:related} compares our contributions with related work. \textcolor{black}{ Sec.~\ref{sec:reusability} describes how we envision our overall approach being replicated for other regulations and document types.} Sec.~\ref{sec:conclusion} concludes the paper.

\vspace{2em}

\section{Background} \label{sec:background}

\textcolor{black}{In this section, we first briefly introduce GDPR. We then summarize the necessary background related to our technical approach.} 

\subsection{GDPR} 
GDPR \cite{EU2018} is a complex regulation comprised of 173 recitals and 99 articles divided into 11 chapters. 
GDPR applies primarily to organizations within \textcolor{black}{Europe}. However, the regulation may also apply to organizations outside \textcolor{black}{Europe}, e.g., when these organizations offer goods or services to, or monitor individuals in \textcolor{black}{Europe}.
If an organization is subject to GDPR, it has to identify itself as either a data controller or data processor. A controller determines the purpose and means of processing, whereas a processor acts on the instructions of the controller.
The responsibilities of a given organization under GDPR vary depending on \hbox{whether it is a processor or a controller.}
Processors notably have to: (1) implement adequate technical and organizational measures to keep personal data  safe and secure, and, in cases of data breaches, notify the controllers; (2) appoint a statutory data protection officer (if needed) and conduct a formal impact assessment for certain types of high-risk processing; (3) keep records about their data processing; and (4) comply to GDPR restrictions when transferring personal data outside \textcolor{black}{Europe}.
In comparison to processors, controllers are subject to more provisions. In particular, in addition to having to meet the obligations mentioned above, controllers have to: (1)~adhere to six core personal data processing principles, namely, fair and lawful processing, purpose limitation, data minimization, data accuracy, storage limitation, and data security; (2)~keep identifiable individuals informed about how their personal data will be used; and (3)~preserve 
the individual rights envisaged by GDPR, e.g., the right to be forgotten \hbox{and the right to lodge a complaint.} GDPR includes some specific provisions in relation to privacy policies.  \textcolor{black}{Privacy policies play a major role in software development. For example, they refer to how a controller (i.e., the software) should ensure data security, how long the collected data should be stored on the controller database, what the software needs to provide to the user if some specific user rights are in place (e.g., withdraw consent and data erasure), etc. We elaborate the GDPR provisions for privacy policies in Sec.~\ref{sec:qualitative}.}

\subsection{Natural Language Processing} \label{subsec:nlp}

Natural language processing (NLP) is a sub-field of AI, 
\textcolor{black}{
which is used for
automated processing of  natural-language data. Examples of NLP applications include machine translation and information extraction~\cite{hirschberg:15, Jurafsky:09}.} 
\begin{wrapfigure}{r} {0.24\textwidth} 
\includegraphics[width=0.23\textwidth]{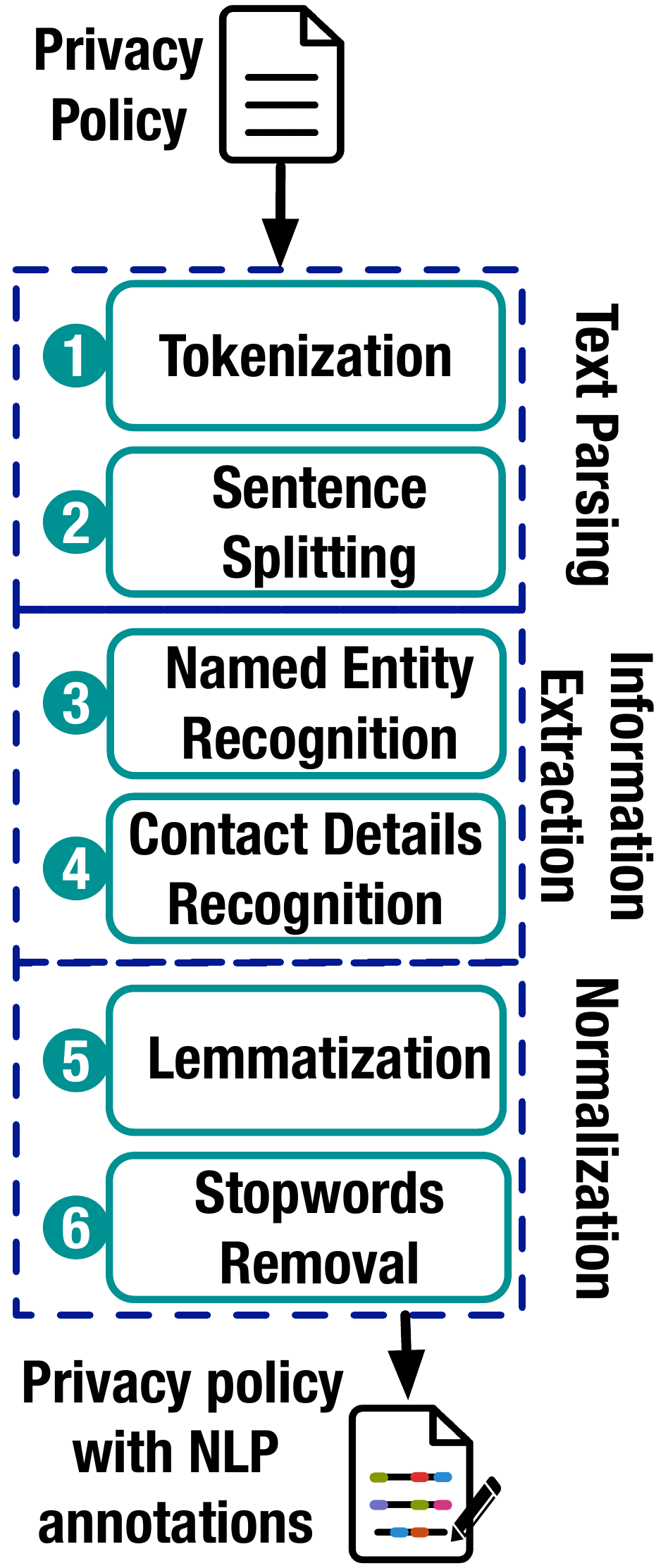}
\caption{NLP Pipeline. } \label{fig:pipeline} 
\end{wrapfigure}
In our work, we apply the NLP pipeline depicted in Fig.~\ref{fig:pipeline}. The pipeline combines six consecutive NLP modules divided in three categories.

The first category is aimed at parsing the text of a given privacy policy. This category includes \emph{Tokenization} for separating out the words and punctuation marks from the running text and \emph{Sentence Splitting} for decomposing the text into coherent sentences based on sentence boundary indicators such as periods, question and exclamation marks~\cite{rodrigues:15, Jurafsky:09}. 

The second category in the pipeline is concerned with extracting information from the text. The first step uses the \emph{Named Entity Recognition (NER)}, which is the task of marking the mentions of named entities in a given text 
with their types~\cite{mikheev:99}, e.g., a country name like ``Luxembourg'' will be annotated with the type \textit{location}. 
The entity types, in our work, are limited to \textit{location} and \textit{organization} since these two are expected to appear in a privacy policy. In addition to the NER module, we use regular expressions~\cite{friedl:06} to recognize the contact details that are mentioned in the input privacy policy, namely email and postal addresses, telephone numbers and websites. For example, the email address ``info@hikari.jp'' will be recognized as \textit{email}.

The last category involves normalizing the text. In particular, different words in a text can be mapped 
to a single root 
form using the \emph{Lemmatization} module, e.g., the words ``deletion'', ``deleted'' and ``delete'' will be lemmatized to the word ``delete''. Finally, we use the \emph{Stopwords Removal} to remove stopwords, i.e., very frequent 
words 
such as prepositions (e.g., ``in'') and articles (e.g., ``a'' and ``the'').  
Applying the NLP pipeline above results in adding various annotations to the input privacy policy.  

\subsection{Machine Learning}  \label{subsec:ml}

Machine learning (ML) is another sub-field of AI which describes the automated learning methods used for finding meaningful patterns in data~\cite{jordan:15, aggarwal:18, Witten:16}. 
Supervised ML assumes that training examples (input) are  provided with their labels (output). Using these training examples, the machine then learns to predict the output of unseen examples. We will refer to the input and its associated output value as a \textit{classification instance}.
Text classification (also known as text categorization) is supervised learning for categorizing the text into a set of predefined groups~\cite{aggarwal:18}, e.g., classifying the text of an email into 
\textit{spam} and \textit{not spam}. 
 
In this work, we focus on \textit{multiclass multilabel classification}. Multiclass classification is to classify the input examples into three or more predefined classes. A classical example in the ML literature is classifying an iris flower, given its sepal length and width and petal length and width, into one of the three possible types \textit{setosa}, \textit{versicolor}, or \textit{virginica}~\cite{Witten:16}. 
Multilabel classification means that the same input example may belong to multiple classes, e.g., classifying movies into one or more genres based on the plot summary, where a movie can belong to \textit{comedy} and \textit{action} at the same time.
The multi-label classification problem is often simplified into multiple binary classification problems~\cite{aggarwal:18}. A binary classification is a specific case of the multi-class classification with only two target classes. For example, a movie can be classified into genres using multiple learning algorithms, such that each learner predicts whether the movie is from a specific genre (e.g., \textit{comedy}) or not from that genre (e.g., \textit{not comedy}), and so on for the other genres. 
 
\subsection{Vector-space Representation of Text}  \label{subsec:embeddings}
\textcolor{black}{\textit{Vectorization} is a prerequisite step to text classification where the text has to be transformed into a set of feature vectors for describing the text under the different pre-defined classes~\cite{schutze:08, aggarwal:18}. Each 
classification instance is represented by a feature vector.
These features can be either manually crafted (e.g., the presence of first person pronouns like ``we'') or automatically generated using the words in the text. 
There are several models to perform vectorization, e.g., BoW (bags of words), TF/IDF (term frequency/inverse document frequency) and word embeddings. 
In our work, we use word embeddings. In particular, we utilize pre-trained word vectors from the GloVe model~\cite{pennington:14}. }

\textcolor{black}{\textit{Word embeddings} are representations of words as dense numerical vectors that capture the syntactic and semantic regularities~\cite{mikolov:13c, pennington:14, Levy:14}. Deriving these representations is based on the distributional hypothesis of Harris~\cite{Harris:54} which states that semantically-related words appear in similar contexts. Therefore, vectors that are close to each other in the vector space should represent words that are similar, e.g., the vector representing the word ``frog'' should be close to vectors of similar words such as ``toad'' and ``lizard''~\cite{pennington:14}. Regularities are observed in the linear relations between word pairs. For example, if a word $w$ is represented by the vector $\vec{w}$, then we observe the plural relation: $\vec{cat} - \vec{cats} \approx \vec{apple} - \vec{apples}$. }

\textcolor{black}{Out of the available methods for learning word embeddings, we use the pre-trained  GloVe embeddings~\cite{pennington:14}. Pre-trained models are used to improve a range of NLP tasks~\cite{joshi:16, Mu:18, yu:17, ghosh:15, iyyer:15, qi:18, kim:14, chen:14}. 
In modern NLP, pre-trained word embeddings perform better than those learned from scratch~\cite{turian:10}.
Compared to newer technologies for generating text representations like ELMo~\cite{Peters:18}, OpenAI GPT~\cite{Radford:18} and BERT~\cite{Devlin:18}, GloVe provides context-independent word embeddings (i.e., one-to-one mapping between the words and their vectors) that can be directly used off-the-shelf. Despite being powerful, context-aware representations generated by (for example) BERT come with the cost of an extra step for training, or fine-tuning. Moreover, the GloVe pre-trained model achieves good results on NLP downstream tasks~\cite{ethayarajh:19}. 
Compared to word2vec~\cite{mikolov:13c} and fasttext~\cite{bojanowski:17}, which also provide pre-trained word vectors, GloVe learns words representations using both local and global context to better capture the semantics of words~\cite{huang:12}. Global context is used to enrich the words representations by considering the co-occurrence counts of the words in a large corpus. }


\begin{figure*}[ht!]
\centering
\includegraphics[width=\linewidth] {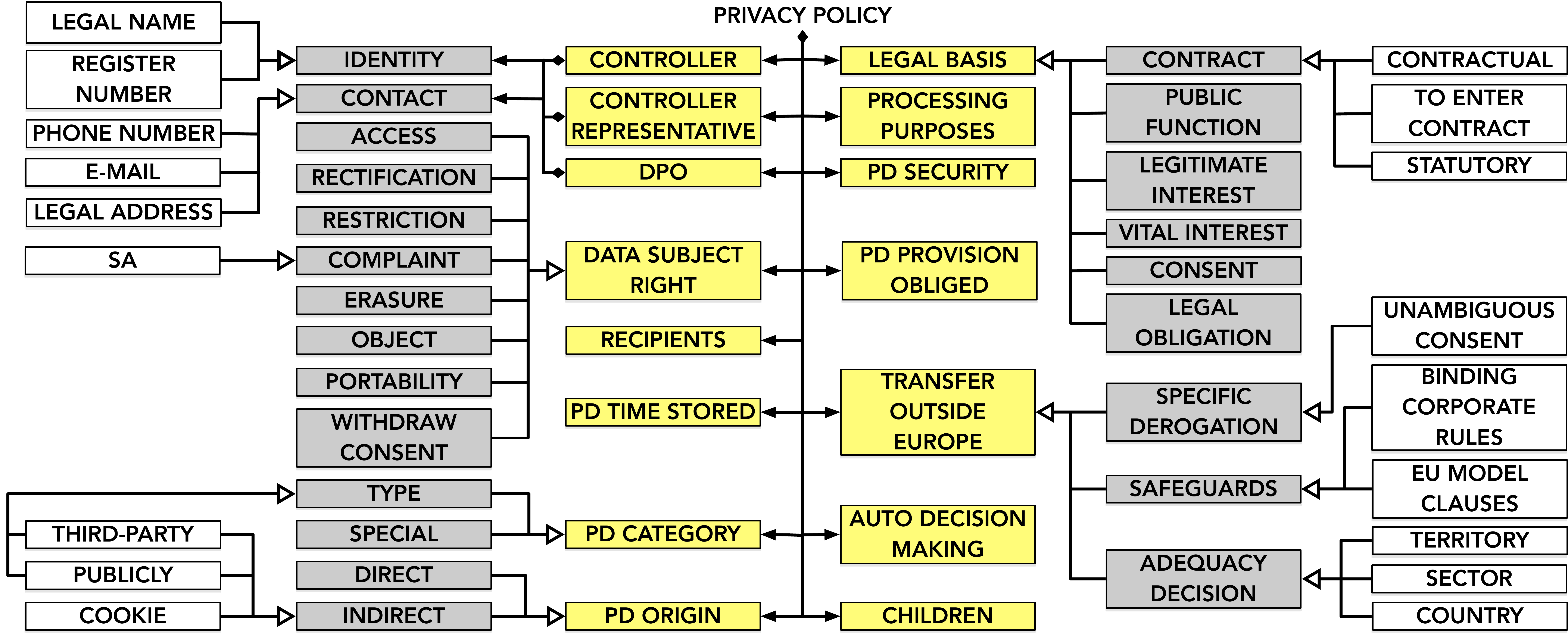}
\caption{Conceptual Model of Privacy-Policy Metadata.} \label{fig:metadata}
\vspace*{-1em}
\end{figure*}

\textcolor{black}{The GloVe pre-trained model, used in our work, uses 100-dimensional vectors generated by training on extensive text corpora from Wikipedia and the web.  
To illustrate, consider the text segment ``Hikari Bank Privacy Policy'' in Fig.~\ref{fig:annotations}. Using  pre-trained word embeddings, each word is represented as a 100-dimensional vector, e.g., ``hikari'' is represented as $[0.42192,0.41032,0.23888,\ldots]_{100}$. Computing the vector representation of this text segment can then be performed by combining the word embeddings in the segment through different mathematical operations including summation and (simple or weighted) averaging~\cite{Blacoe:12, zhu:18, Wieting:15}. In our approach, we use simple averaging because it proved to be effective in text-similarity-related tasks. }

\section{A Conceptual Model of Privacy-Policy Metadata (RQ1)} \label{sec:qualitative}

\begin{figure}
\centering
\includegraphics[width=\linewidth] {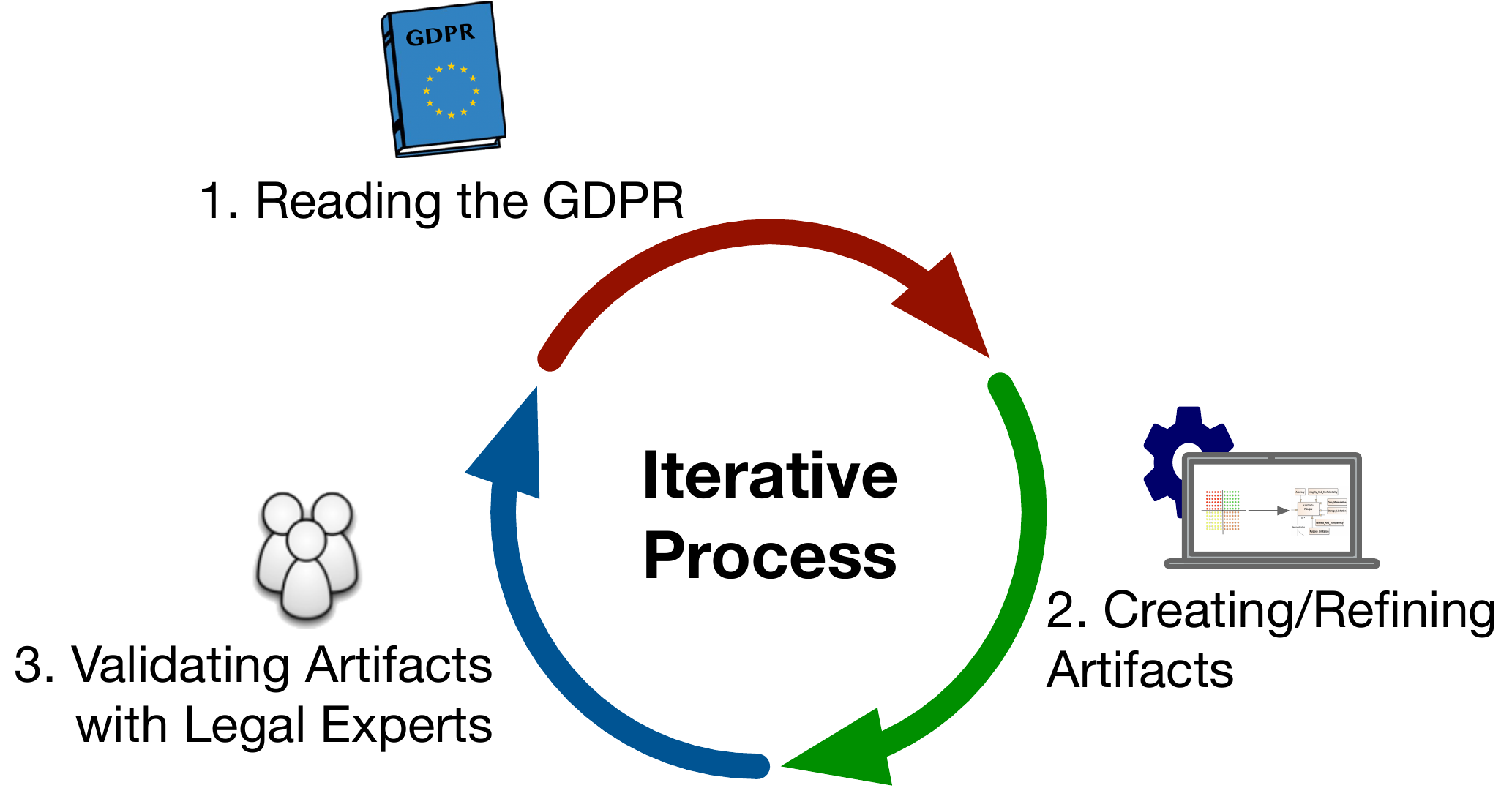}
\caption{Iterative Process.} 
\label{fig:iterativeProcess}
\end{figure}

In this section, we present the following artifacts to answer RQ1: (1) a conceptual model specifying, in a comprehensive manner, the metadata types pertinent to GDPR privacy policies; and (2) a glossary defining all necessary terms to better understand the conceptual model with traceability to GDPR articles.
The conceptual model (artifact 1) is shown in Fig.~\ref{fig:metadata} and an excerpt of the glossary (artifact 2) is presented in Table~\ref{tab:glossary}. The complete glossary is provided as an online annex \cite{ADs_Traceability}.
The above artifacts were built using an iterative and incremental method following three main steps (see Fig. \ref{fig:iterativeProcess}): (1) reading the articles of GDPR that address privacy policies, (2) creating and refining the artifacts introduced above, and (3) validating these artifacts with legal experts. Building the artifacts took four iterations with each iteration requiring, on average, one month. We had several face-to-face and off-line validation sessions with legal experts. The sessions, which lasted between two and three hours each, collectively added up to approximately 30 hours.

\textcolor{black}{
We conducted our validation sessions with three legal experts, namely (a) a senior lawyer with more than 30 years of experience in European and international laws; \hbox{(b) a mid-career} lawyer with more than 10 years of experience in law with a focus on the data protection and financial domains; and (c) an IT professional with more than 10 years of experience in the legal domain. 
Each validation session was attended by at least two legal experts. The discussions continued until the experts in attendance agreed that the model correctly reflected their interpretation of GDPR. We observed that the differing viewpoints and thus the deliberations between the legal experts centered primarily around the specializations in the conceptual model (e.g., the sub-metadata types of \textsc{Legal Basis.Contract}) and about how different metadata types should be inter-related (e.g., how \textsc{PD Origin.Indirect} is related to \textsc{PD Category.Type}).
}

\begin{table}[t]
\caption{Glossary Excerpt. }\label{tab:glossary}
  \centering
  \begin{threeparttable}[t]
  \begin{tabularx}{0.48\textwidth}{@{} p{0.2\textwidth}  @{\hskip 0.2em} p{0.28\textwidth} }
   
   Metadata (\textbf{Reference}\tnote{1}) & Description \\ 
   \toprule
    \textsc{Controller} (\textbf{Art. 13/14(f)}) & A natural or legal person, public authority, agency or any other body which, alone or jointly with others, determines the purposes and means of the processing of personal data where the purposes and means of such processing are determined by national or EU laws or regulations, the controller or the specific criteria for its nomination may be provided by national or EU law.\\
    \midrule
   \textsc{Identity} (\textbf{Art. 13/14(f)}) & The legal name of the company/organization. \\ 
   \midrule
    \textsc{Contact} (\textbf{Art. 13/14(f)}) & The method(s) with which the company/organization can be contacted. \\
    \midrule
    \textsc{Controller} \par \textsc{Representative} \par (\textbf{Art. 13/14(f)})& A natural or legal person established in the Union who is designated by the controller. \\
    \midrule
    \textsc{Data Protection} \par \textsc{Officer} (\textsc{DPO}) \par (\textbf{Art. 13/14(f)}) & 
    The one who is responsible for overseeing data protection strategy and implementation to ensure compliance with GDPR requirements. \\ 
    \midrule
    \textsc{Processing} (\textbf{Art. 13/14(f)}) & Any operation performed on personal data, whether or not by automated means, including collection, recording, organization, structuring, storage, adaptation or alteration, retrieval, consultation, use, disclosure by transmission, dissemination or otherwise making available, alignment or combination, restriction, erasure or destruction. \\
    \midrule
    \textsc{Personal Data (PD)} \par  (\textbf{Art.5(f)}) & Any information related to an identified or identifiable natural person. \\
    \midrule
    \textsc{Provision} (\textbf{Art. 14(f)}) & The action of providing something (i.e., personal data) for use (i.e., to be processed). \\ 
    \midrule
    \textsc{PD Origin} (\textbf{Art.~14.2(f)}) & From which source the personal data originates (i.e., direct or indirect), and if applicable, whether it came from a publicly and/or third-party and/or cookie sources. \\
    \midrule
    \textsc{Indirect} (\textbf{Art. 14}) & When the personal data are not obtained from the data subject. \\ 
    \midrule
    \textsc{Third Party} (\textbf{Art. 14}) & When the personal data are obtained from organisations external to the data controller. \\
    \midrule
    \textsc{Publicly} (\textbf{Art. 14}) & When the personal data are obtained from public sources (i.e., from a public website). \\ 
    \midrule
    \textsc{Profiling} (\textbf{Art. 4(f)}) & To analyze or predict aspects concerning a natural person’s performance at work. \\ 
    \bottomrule
 \end{tabularx}
  \begin{tablenotes}
     \item[1] GDPR-related articles
   \end{tablenotes}
 \end{threeparttable}
 
\end{table}

Initially, as suggested by our collaborating legal experts from Linklaters, we analyzed Art(icles) 13 and 14 of GDPR, i.e., the main GDPR articles targeting  privacy policies. From these two articles, we extracted important concepts to create the metadata and the dependencies between them.
Art.~13 focuses on personal data collected directly from a data subject (e.g., filling an online form or an interview), whereas Art.~14 focuses on personal data obtained indirectly from a data subject (e.g., obtained from a public website or public list). 
We observe that Art. 13.2(e) {\textit{(whether the provision of personal data is a statutory or contractual requirement, or a requirement necessary to enter into a contract, as well as whether the data subject is obliged to provide the personal data and of the possible consequences of failure to provide such data)}} is related to the direct collection of personal data, while Art. 14.2(f) {\textit{(from which source the personal data originate, and if applicable, whether it came from publicly accessible sources)}} deals with indirect collection. These observations were taken into consideration while building the two artifacts discussed above. Starting from Art. 13 and 14, as per the recommendation of legal experts, we also examined Art. 6, 9, 21, 37, 46, 47, 49, 55, and 56 by doing a snowball sampling from the cross-references in Art 13 and 14.

\begin{figure}
\centering
\includegraphics[width=\linewidth] {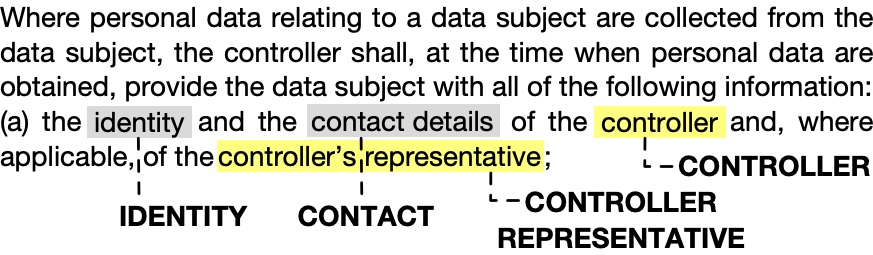}
\caption{Example of Coding in the Context of GDPR.} \label{fig:Art13example}
\vspace*{-1em}
\end{figure}

\begin{wrapfigure}{r} {0.3\textwidth} 
\includegraphics[width=0.3\textwidth]{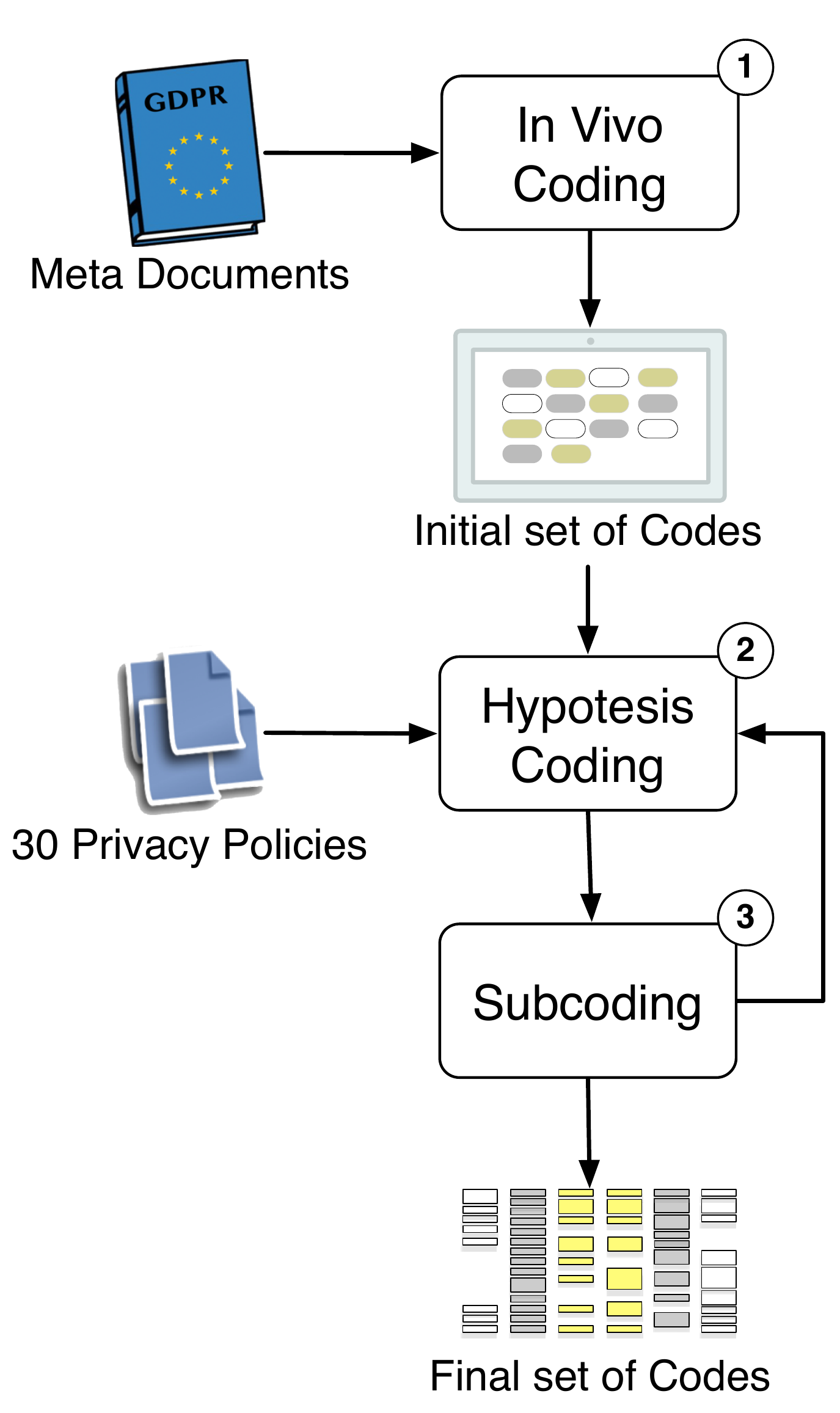}
\caption{Coding methodology.} \label{fig:coding} 
\vspace*{-.7em}
\end{wrapfigure}

Fig.~\ref{fig:Art13example} illustrates an excerpt of Art. 13 from which we have inferred the hierarchical representation of  
four metadata types: \textsc{Controller},  \textsc{Controller Representative}, and their descendants \textsc{Identity} and \textsc{Contact}.
These metadata types refer to four distinct concepts: (1) the identity of the data controller \textsc{(Controller$.$Identity)}, (2) the contact details of the data controller \textsc{(Controller$.$Contact)}, (3) the identity of the data controller's representative \textsc{(Controller Representative.Identity)}, and (4) the contact details of the data controller's representative \textsc{(Controller Representative$.$Contact)}. The metadata types \textsc{Identity} and \textsc{Contact} were ultimately specialized with the inclusion of other sub-metadata types. The former, with \textsc{Legal Name} and \textsc{Register Number} metadata types and the latter with \textsc{Email}, \textsc{Legal Address} and \textsc{Phone Number}.
Our conceptual model (depicted in Fig.~\ref{fig:metadata}) is organized into three hierarchical levels: \textbf{level-1}, shaded yellow, \textbf{level-2}, shaded grey, and \textbf{level-3}, shaded white. The colors were introduced to make the model more readable to annotators and legal experts. \textcolor{black}{As presented in Fig.~\ref{fig:coding}, the methodology we used for identifying the metadata types from GDPR and building the conceptual model involved three types of coding:  \emph{in-vivo coding, hypothesis coding} and \emph{subcoding}~\cite{Saldana2016}.}

\textcolor{black}{\textbf{1. In-vivo coding:} we use this type of coding to identify the core concepts in GDPR and create an initial set of codes. In-vivo coding emphasizes the actual words in the text -- in our case the text of GDPR  -- in order to create codes. The in-vivo approach allowed us to derive the names of the metadata types directly from the text of GDPR (i.e., the meta documents). Those metadata types are then used to characterize GDPR-related text found in privacy policies.
A metadata type, representing a code, is a short phrase that symbolically assigns a summative, salient, essence-capturing, and/or evocative attribute to a particular text in a given privacy policy~\cite{Saldana2016}. For example, the metadata type \textsc{Controller} in Fig. \ref{fig:Art13example} refers to the text in a given privacy policy that discusses a natural or legal person, public authority,  agency  or  any  other  body  which, alone  or  jointly  with  others,  determines the purposes and means of personal data processing (see Art. 13.1(a) of GDPR). 
}

\textcolor{black}{\textbf{2. Hypothesis coding:} this type of coding refers to the application of a predetermined set of codes to qualitative data in order to assess researcher-generated hypotheses. The codes are developed from a prediction -- in our case, the initial set of codes identified from GDPR with in-vivo coding -- about what one would find in the actual data -- in our case, privacy policies -- before the data was collected and analyzed.
Usually, the application of this coding methodology can range from simple frequency counts to more complex multivariate analyses. In our context, we are interested in the presence or absence of metadata types in a given privacy policy in order to check its completeness against GDPR. 
In particular, with the help of legal experts, we manually applied this initial set of codes (obtained with in-vivo coding) via hypothesis coding over 30 privacy policies (a subset of our training set) in order to ensure that our in-vivo codes are sufficient and at the right level of abstraction. }

\textcolor{black}{While applying hypothesis coding to the privacy policies, for each metadata type, we collected the keywords that made us decide to associate a given sentence with a specific metadata. For example, the combination of keywords \textit{"right to access"} was extracted from sentence number \textbf{27} of Fig.~\ref{fig:annotations} and included in the list of of keywords associated with \textsc{Data Subject Rights.Access}. At the end of hypothesis coding, we obtained a list of keywords for each metadata type as shown in Fig.~\ref{fig:metadata}.}

\textcolor{black}{\textbf{3. Subcoding:} in addition to hypothesis coding, we also use subcoding, which refers to sub-codes as a second-order tag assigned after a primary code, in order to enrich our metadata types in terms of specificity. For example, in Fig. \ref{fig:metadata}, the metadata type \textsc{PD Origin} (in yellow) is specialized into two sub-metadata types: \textsc{Direct} and \textsc{Indirect} (in gray). Then, \textsc{Indirect} is further specialized into: \textsc{Third-Party}, \textsc{Publicly} and \textsc{Cookie} (in white). The use of subcoding ultimately contributed to the final set of codes represented by the conceptual model of Fig.~\ref{fig:metadata}.}

Based on our interpretation and understanding of GDPR articles, we created an initial version of the metadata conceptual model along with their definitions. We kept track of GDPR articles to ensure traceability in our glossary (artifact 2). Table~\ref{tab:glossary} presents an excerpt of our glossary.

These (interim) artifacts were then presented to legal experts for feedback. 
In addition to pointing out issues and omissions, the experts were encouraged to bring to our attention any GDPR article or external documentation/information needed to be considered in the context of privacy policies.The feedback obtained from legal experts was, by and large, concerned with information that was not explicitly included in GDPR (e.g., the European Working Party~\cite{WP2018}). For example, Art. 13.1(f) states 
that "{\textit{the controller intends to transfer personal data to a third country or international organization and the existence or absence of an adequacy decision by the Commission, or [...] appropriate or suitable safeguards [...]}}". This article is addressed in Fig.~\ref{fig:metadata} by the metadata type \textsc{Transfer Outside Europe$.$Adequacy Decision}. 
In response to the legal experts' feedback and by following the external source\footnote{EU Adequacy Decisions -- https://bit.ly/38ciwPU (January 2021)} recommended by them, we created the sub-metadata types of \textsc{Adequacy Decision} that are not discussed in GDPR. In particular, three sub-metadata types were added to the conceptual model in Fig.~\ref{fig:metadata}, namely, \textsc{Territory}, \textsc{Sector}, and \textsc{Country}. These metadata types refer to the adequacy decisions between the EU and a territory (e.g., Andorra, the Bailiwick of Jersey, etc.), specific sectors (e.g., the commercial organizations from Canada, Argentina, etc.), and a country (i.e., Japan, New Zealand, etc.), respectively. In the same manner, we used another external source\footnote{EU Standard Contractual Clauses -- https://bit.ly/3nd6JFt (January 2021)} to create the level-3 sub-metadata type \textsc{EU Model Clauses}. 

Once the conceptual model converged to a stable state, we put together a general report including the conceptual model and the glossary table. The conceptual modeling step terminated when the general report was approved by legal experts. The final version of the conceptual model, with a total of 56 metadata types (see Fig.~\ref{fig:metadata}), along with a complete glossary table of 60 entries, are provided in an online annex~\cite{ADs_Traceability}.


\section{Completeness Checking Criteria for Privacy Policies (RQ2)} \label{sec:criteria}

In this section, we answer RQ2 by presenting the criteria we use to check the completeness of privacy policies according to GDPR. In particular, we discuss our method for creating a set of 23 criteria for checking the completeness of privacy policies by analyzing GDPR articles. In order to identify these criteria, we used an iterative three-step method similar to the one we used to create the other two artifacts mentioned in Sec.~\ref{sec:qualitative} (see Fig.\ref{fig:iterativeProcess}). We obtained the final set of completeness criteria in six iterations, with each iteration requiring, on average, 15 days. During this process, we combined bi-weekly face-to-face validation sessions and off-line interactions with legal experts. The face-to-face sessions, which lasted between 2 to 3 hours each, collectively added up to approximately 15 hours, plus an additional five hours for off-line interactions.

\subsection{Transforming GDPR Articles into Criteria}\label{subsec:CriteriaCreation} 
The complete set of criteria discussed in this section uses the metadata types identified in the conceptual model of Fig.~\ref{fig:metadata}. 
\textcolor{black}{We note that some metadata types are \textit{inter-dependent}, meaning that 
the presence of a metadata type requires the presence of another metadata type. For example, if a privacy policy requires individuals to provide consent for collecting their personal data, then the policy shall also allow individuals to withdraw their consent, i.e., \textsc{Legal Basis.Consent}  and \textsc{Data Subject Right.Withdraw Consent} are inter-dependent. 
} 
Most of the criteria were extracted from the same GDPR articles from which the metadata types were also identified (see the external online annex \cite{ADs_Traceability} for criteria traceability to the GDPR articles). Based on our interpretation and understanding of these GDPR articles, we identified an initial set of criteria that we formulated as pseudo-code. Each pseudo-code statement is composed of two main parts: 1) a \textit{precondition} (if any) about the identification of one or more metadata types in a privacy policy or other GDPR-related conditions proposed by the legal experts, and 2) a \textit{postcondition} asserting the identification of one or more metadata types (different from the one(s) in the \textit{precondition}) in a privacy policy. We use the following template [\textit{precondition}], <\textit{postcondition}> and show below examples of criteria written in pseudo-code; these are derived from the excerpt of Art. 13.1(a) shown in Fig.~\ref{fig:Art13example}:
\begin{itemize}
    \item[C1] [ ], <\textsc{Controller.Identity.}\{\textsc{Register Number} \textit{or} \textsc{Legal Name}\} must be identified>. 
    \item[C2] [ ], <\textsc{Contact.}\{\textsc{Email} or \textsc{Phone} or \textsc{Legal Address}\} must be identified>.
    \item[C3] [\textit{if} \textsc{Controller} is located outside of Europe], <\textit{then} \textsc{Controller Representative.Identity.}\{\textsc{Register Number} or \textsc{Legal Name}\} must be identified>.
    \item[C4] [\textit{if} \textsc{Controller} is located outside of Europe], <\textit{then} \textsc{Controller Representative.Contact.}\{\textsc{Email} or \textsc{Phone} or \textsc{Legal Address}\} must be identified>. 
\end{itemize}

In this step, we transform the text of the relevant GDPR provisions into completeness criteria. For example, considering Fig.~\ref{fig:Art13example}, the word \textit{shall} 
is translated into a mandatory requirement for including  
the \textsc{Controller.Identity} (C1) and \textsc{Controller.Contact} details (C2). On the other hand, the combination of the words \textit{where} and \textit{applicable}
suggests that a given criterion should be enforced only if certain precondition(s) are met: the \textsc{Controller Representative.Identity} (C3) and \textsc{Controller Representative.Contact} (C4) need to be checked only if the \textsc{Controller} is located outside of \textcolor{black}{Europe}.

While defining the criteria from the GDPR articles, we realized that some of them should not always be checked. 
Articles 13.1(a,e,f), 13.2(e), 14.1(a,d,e,f), and 14.2(f) are GDPR articles that apply to privacy policies only in specific situations. After reviewing these articles with legal experts, they asked us to create a questionnaire that would help them specify, under various situations, the exact content of a given privacy policy for completeness checking. 
\textcolor{black}{The person who should ideally provide answers to the questionnaire should have expertise in the legal domain as well as extensive knowledge about the company for which the privacy policy analysis is being performed.  
For example, to determine whether C3 and C4 above should be checked, it is important to know beforehand from the questionnaire that the \textsc{Controller} is located outside Europe.} 

The questionnaire contains a set of critical questions whose answers depend on context and are often left tacit in privacy policies. Nevertheless, these answers carry important implications on what needs to be explicitly covered in privacy policies, and hence on completeness checking. The questionnaire includes the following six questions:
\begin{itemize}
    \item[\textbf{Q1}] Who is the \textsc{Controller} in charge of data processing? \textit{Write name}.
    \item[\textbf{Q2}] Do you plan to transfer the collected personal data outside \textcolor{black}{Europe}? \textit{Yes/No}.
    \item[\textbf{Q3}] Will there be other recipients of the collected personal data besides you? \textit{Yes/No}.
    \item[\textbf{Q4}] What is the core of your activities? 
    \par $\square$ The processing of personal data is carried out by a public authority or body (except for courts acting in their judicial capacity).
    \par $\square$ The processing of operations which, by nature, scope and/or purposes, require regular and systematic monitoring of data subjects on a large scale. 
    \par $\square$ The processing, on a large scale, of personal data relating to sensitive categories (e.g., racial or ethnic origin, political opinions, or religious or philosophical beliefs) or to criminal convictions and offenses. 
    \item [\textbf{Q5}] Where will the activities carried out by your organization take place? 
        \par $\ocircle$ \textit{Inside \textcolor{black}{Europe}}
        \par $\ocircle$ \textit{Outside \textcolor{black}{Europe}} -- if selected, then write the name of  \par \hfill \fontsize{8}{10}{\textsc{Controller Representative}:} \ldots \ldots 
    \item[\textbf{Q6}] How will the personal data of the data subject be collected? \hfill $\ocircle$ \textsc{Direct} \hfill $\ocircle$ \textsc{Indirect} \hfill $\ocircle$ \textit{Both}
\end{itemize}

Question \textbf{Q1} is not intended to trigger the checking of any criterion. This first question is used to facilitate the identification of the metadata type \textsc{Controller.Identity}.
\newline The main objective of the remaining questions is to determine whether some 
context-related criteria should be checked.
In particular, each of the other five questions (\textbf{Q2-Q6}) triggers the search for one or more metadata types. This leads to checking some specific criteria. A positive answer to question Q2 triggers the verification of criteria C10 -- C14. 
If the answer to Q3 is yes, then criterion C19 is verified. Similarly, Q4 will trigger the verification of criterion C23, if any of the optional answers to this question is checked. The answer to  question Q5 activates the verification of criteria C3 and C4, if the activities carried out by the \textsc{Controller} take place outside Europe. 
Finally, answering Q6 as ``\textsc{Direct}'' activates checking criterion C22;       ``\textsc{Indirect}'' activates checking criteria C15 -- C18; and ``\textit{Both}'' requires checking all the above criteria (i.e., C15 -- 18 and C22).

\newcolumntype{P}[1]{>{\centering\arraybackslash}p{#1}}

\begin{table*}[t!]
\caption{Completeness Criteria according to GDPR. }\label{tab:criteria_improved}

  \centering
  \begin{threeparttable}[t]
  \begin{tabularx}{0.98\textwidth}{@{} P{0.05\textwidth}
    @{\hskip 0.2em} p{0.28\textwidth} @{\hskip 0.3em} p{0.65\textwidth}}
   
    \multirow{2}{*}{ID} & \multicolumn{2}{c}{Criteria} \\
    & \multicolumn{1}{c}{Precondition\tnote{1}} & \multicolumn{1}{c}{Postcondition\tnote{2}} \\
    \toprule
    \colorbox{red!25}{C1} & \multicolumn{1}{c}{-} &  \textsc{Controller.Identity} \\
    \midrule
    
    \colorbox{red!25}{C2} & \multicolumn{1}{c}{-} & \textsc{Controller.Contact}.\{\textsc{Legal Address, Email}, \textit{or} \textsc{Phone Number}\} 
    \\
    \midrule
    
    \colorbox{red!25}{C3} & \textbf{A5} is a country \textit{outside the EU} & \textsc{Controller} \textsc{Representative}\textsc{.Identity}\\
    \midrule
    
    \colorbox{red!25}{C4} & \textbf{A5} is a country \textit{outside the EU} & \textsc{Controller} \textsc{Representative} \textsc{.Contact}.\{\textsc{Legal Address},  \textsc{Email}, \textit{or} \textsc{Phone Number\}} 
    \\
    \midrule
    
    \colorbox{red!25}{C5} & \multicolumn{1}{c}{-} & \textsc{Data Subject Right}.\{\textsc{Access, Complaint,} \textsc{Rectification}, \textit{and} \textsc{Restriction}\} \\
    \midrule
    
    \colorbox{orange!25}{C6} & \textsc{Data Subject Right}\textsc{.Complaint} & \textsc{Data Subject Right}\textsc{.Complaint.SA} \\
    \midrule
    
    \colorbox{red!25}{C7} & \textsc{Legal Basis.Contract} & \textsc{Data Subject Right}\textsc{.Portability} \\
    \midrule
    
    \colorbox{red!25}{C8} & \textsc{Legal Basis}  .\{\textsc{Legitimate Interest} \par \textit{or}  \textsc{Public Function}\} & \textsc{Data Subject Right.Object} \\
    \midrule
    
    \colorbox{red!25}{C9} & \textsc{Legal Basis.Consent} & \textsc{Data Subject Right}.\{\textsc{Erasure, Object, Portability,} \textit{and} \textsc{Withdraw Consent\}} \\
    \midrule
    
    \colorbox{red!25}{C10} & \textbf{A2} is \textit{Yes} & \textsc{Transfer Outside Europe} \\
    \midrule
    
    \colorbox{orange!25}{C11} & \textsc{Transfer Outside Europe} & \textsc{Transfer Outside Europe}.\{\textsc{Adequacy Decision, Safeguards,} \textit{or} \textsc{Specific Derogation}\} \\
    \midrule
    
    \colorbox{orange!25}{C12} & \textsc{Transfer Outside Europe} \par \textsc{.Adequacy Decision} & \textsc{Transfer Outside Europe}.\textsc{Adequacy Decision} .\{\textsc{Country,} \textsc{Sector,} \textit{or} \textsc{Territory}\} \\
    \midrule
    
    \colorbox{orange!25}{C13} & \textsc{Transfer Outside Europe}\par \textsc{.Safeguards} & \textsc{Transfer Outside Europe}\textsc{.Safeguards}.\{\textsc{EU Model Clauses,} \par \textit{or} \textsc{Binding Corporate Rules\}} \\
    \midrule
    
    \colorbox{orange!25}{C14} & \textsc{Transfer Outside Europe}\par\textsc{.Specific Derogation} & \textsc{Transfer Outside Europe}\textsc{.Specific Derogation}\textsc{.Unambiguous Consent} \\
    \midrule
    
    \colorbox{red!25}{C15} & \textbf{A6} is \textsc{Indirect} or \textit{Both} & \textsc{PD Origin.Indirect}\\
    \midrule
    
    \colorbox{orange!25}{C16} & \textsc{PD Origin.Indirect} & \textsc{PD Origin.Indirect}.\{\textsc{Third Party,} \textit{or} \textsc{Publicly}\} \\
    \midrule
    
    \colorbox{red!25}{C17} & \textbf{A6} is \textsc{Indirect} or \textit{Both} & \textsc{PD Category}\\
    \midrule
    
    \colorbox{orange!25}{C18} & \textsc{PD Origin.Indirect}\par.\{\textsc{Third Party,} \textit{or} \textsc{Publicly}\} & \textsc{PD Category.Type}\\
    \midrule
    
    \colorbox{red!25}{C19} & \textbf{A3} is \textit{Yes} & \textsc{Recipients}\\
    \midrule
    
    \colorbox{red!25}{C20} & \multicolumn{1}{c}{-} & \textsc{PD Time Stored}\\
    \midrule
    
    \colorbox{red!25}{C21} & \multicolumn{1}{c}{-} & \textsc{Processing Purposes}\\
    \midrule
    
    \colorbox{red!25}{C22} & \textbf{A6} is \textsc{Direct} or \textit{Both} 
    \textbf{and} \textsc{Legal Basis} \par .\{\textsc{Contract.To Enter Contract}, \par \textit{or} \textsc{Legal Obligation}\} & \textsc{PD Provision Obliged}\\
    \midrule
    
    \colorbox{red!25}{C23} & At least one answer in \textbf{Q4} is selected & \textsc{DPO.Contact.}\{\textsc{Legal Address,} \textsc{Email} \textit{or} \textsc{Phone Number}\}\\
    \bottomrule
 \end{tabularx}
  \begin{tablenotes}
     \item[1] Includes the answers to Q2 -- Q6 (\textbf{A2 -- A6}), or the metadata types that are present in a privacy policy.
     \item[2] Metadata types that \colorbox{red!25}{must}/\colorbox{orange!25}{should} be present.
   \end{tablenotes}
 \end{threeparttable}
 \vspace{-.5em}
\end{table*}

The remaining criteria (C1, C2, C5 -- C9, C20 and C21) are always verified because they refer to metadata types that, according to GDPR, must be present in every privacy policy.

At the end of the first step, we created a table with all the information about the criteria set, including an identifier (ID) for each criterion (first column), preconditions (middle column) and postconditions (last column), as shown in Table \ref{tab:criteria_improved}. Since C1, C2, C5, C20 and C21 are not triggered by any preconditions, they always need to be checked. The rest of the criteria are triggered by some precondition related to answers to questions Q2 -- Q6 (referred to as A2 -- A6) from the questionnaire or the presence of some metadata in the privacy policy.

Table \ref{tab:criteria_improved} presents criteria that \emph{should} be satisfied according to GDPR (ID highlighted in orange) and may lead to a warning, and other criteria that \emph{must} always be satisfied (ID highlighted in red) and may lead to a violation. We further discuss the difference between warnings and violations in the next subsection. 


\begin{figure*}[!t]
\centering
\includegraphics[width=0.97\linewidth] {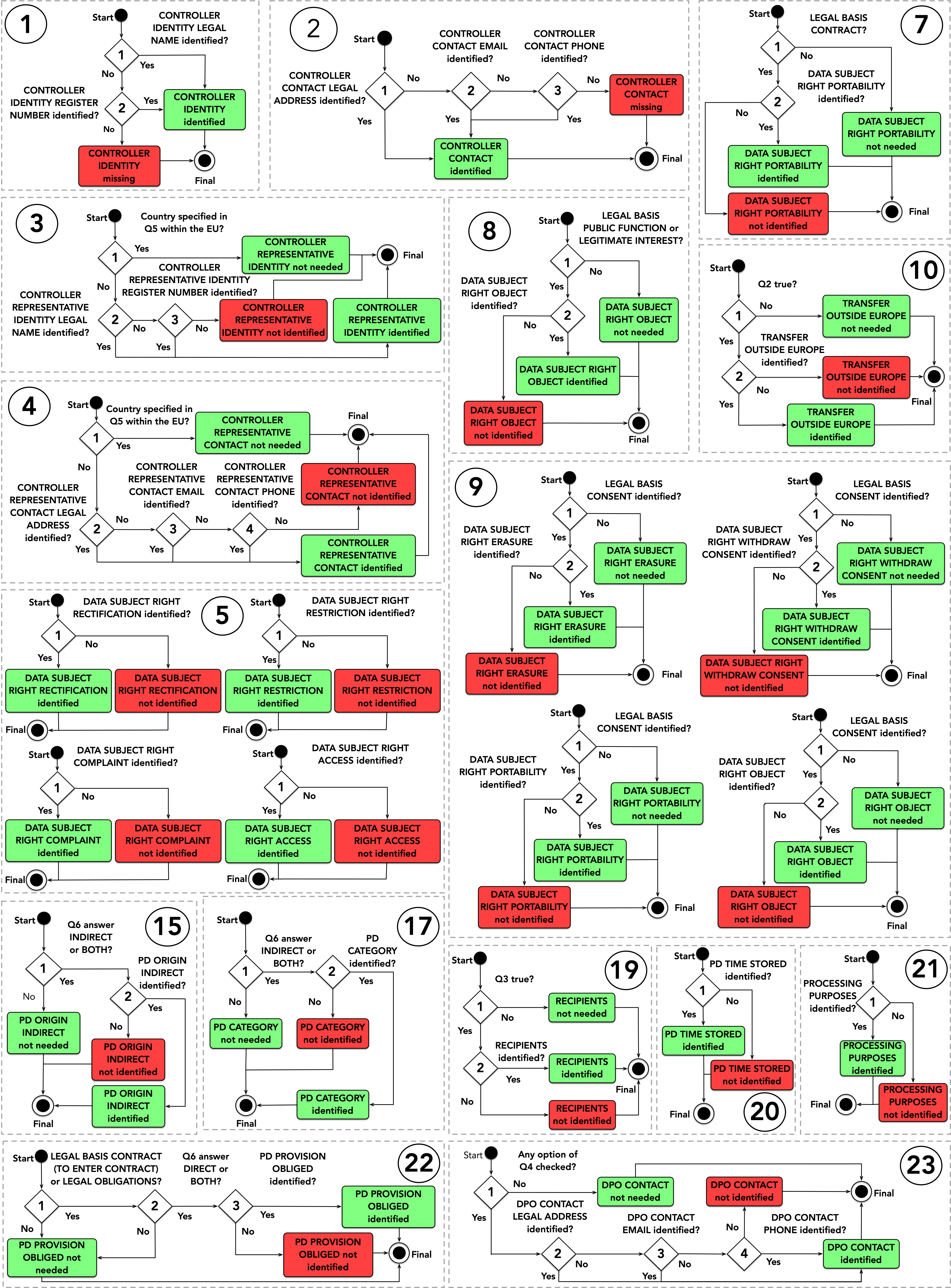}
\caption{GDPR Completeness Criteria Represented as Activity Diagrams (Violations).} \label{fig:Violations}
\end{figure*}

\begin{figure*}[!t]
\centering
\includegraphics[width=0.97\linewidth] {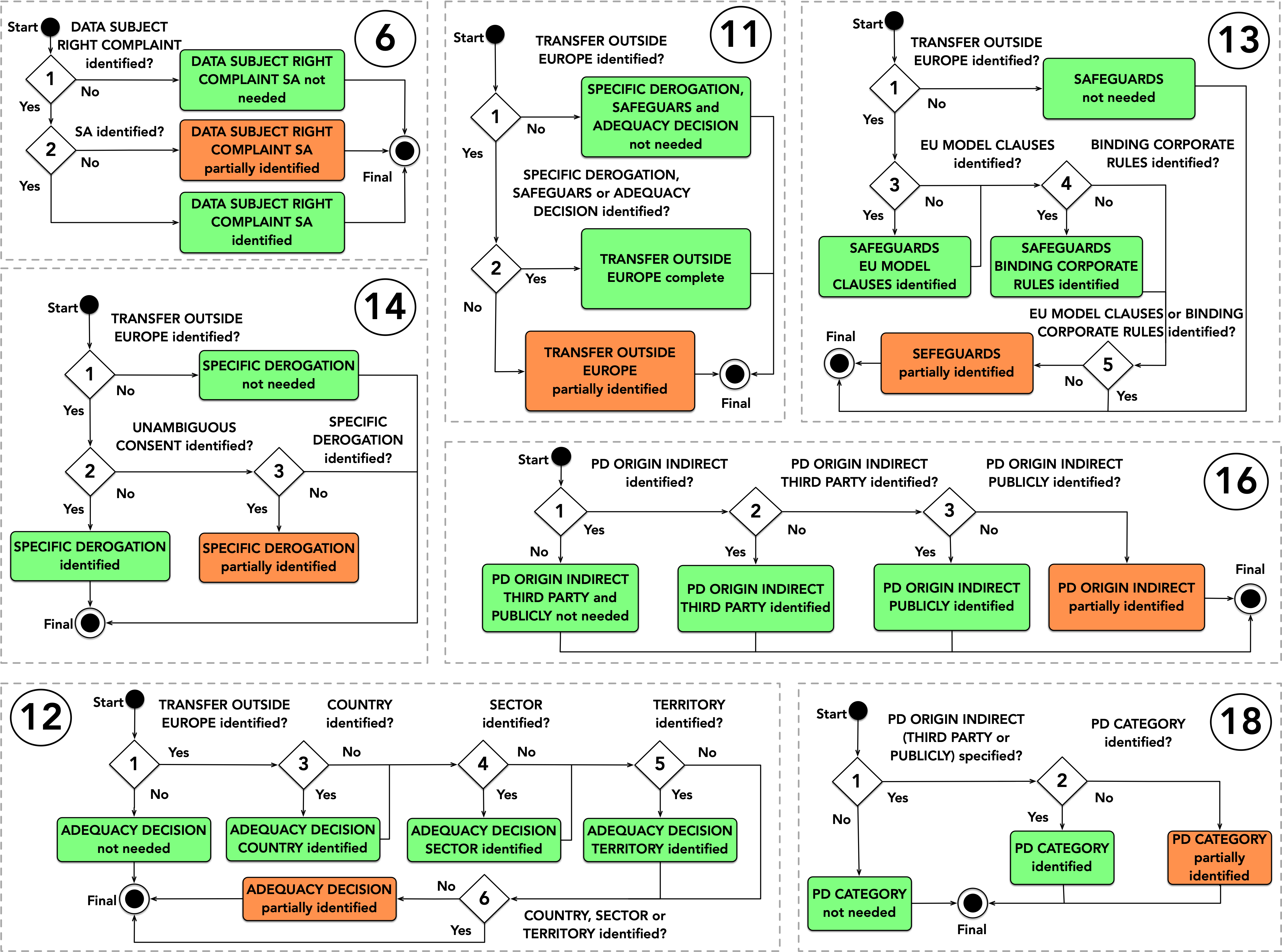}
\caption{GDPR Completeness Criteria Represented as Activity Diagrams (Warnings).} \label{fig:Warning}
\end{figure*}

\subsection{Evaluating the Criteria with Legal Experts} 
To facilitate the validation of the 
criteria presented in Table~\ref{tab:criteria_improved} with legal experts, we decided to capture them as activity diagrams, following the observation by Soltana et al.~\cite{Soltana2018} that legal experts can understand activity diagrams with relative ease given some basic training. With the help of legal experts, we created a final set of 23 criteria to capture the mechanisms necessary to check the completeness of privacy policies according to GDPR. Among the 16 criteria shown in Fig.~\ref{fig:Violations}, C3, C4, C15, C17, C19, and C23 depend on 
the answers to the questionnaire. Fig.~\ref{fig:Violations} and Fig.~\ref{fig:Warning} show the 23 criteria to check the completeness of a privacy policy with respect to the metadata types of the conceptual model presented earlier. Fig.~\ref{fig:Violations} contains every possible violation in privacy policies and Fig.~\ref{fig:Warning} all the possible warnings. 

The completeness criteria in Fig.~\ref{fig:Violations} and Fig.~\ref{fig:Warning} use in general three shapes to represent different types of actions or steps in a process: (1) a circle represents the start and  endpoint, (2) a diamond indicates a decision, and (3) a rectangle stands for an action representing that (3.1) a metadata type was correctly identified or not needed in a privacy policy (in green), (3.2) a mandatory metadata type was entirely missing in a privacy policy (referred to as \textit{violation} -- highlighted in red), and (3.3) a metadata type was only partially identified or, in other words, a metadata type was identified but some related information is missing (referred to as \textit{warning} -- highlighted in orange).  
An \textit{incompleteness issue} is raised when a criterion returns a violation or warning. 
A violation corresponds to a direct breach of GDPR, whereas a warning 
leads to further assessment by the legal expert to  finally decide whether there is a breach of GDPR. 

Below, we illustrate two criteria, C15 and C16, derived from Art. 14.2(f) of the GDPR (see Fig.~\ref{fig:Violations} and Fig.~\ref{fig:Warning}). These criteria check the completeness of a privacy policy with respect to the metadata type \textsc{PD Origin.Indirect}. C15 is meant for identifying  a violation: \begin{itemize}
    \item[\textbf{(1)}] If the answer to \textbf{Q6} is \textsc{Indirect} or \textit{Both} (recall the questionnaire presented in Section~\ref{subsec:CriteriaCreation}), then go to \textbf{(2)}; otherwise 
    \textsc{PD Origin.Indirect} is \emph{not needed}.
    \item[\textbf{(2)}] If the indirect origin of the personal data is mentioned, then \textsc{PD Origin.Indirect} is \textit{identified}; otherwise \textsc{PD Origin.Indirect} is \emph{missing} -- \textbf{Violation}.
\end{itemize}
 Criterion C16 is meant for identifying 
 a warning: 
\begin{itemize}
    \item[\textbf{(1)}] If \textsc{PD Origin.Indirect} is identified in C15, then go to \textbf{(2)}; otherwise 
    \textsc{PD Origin.Indirect} is \emph{not needed}.
    \item[\textbf{(2)}] If the indirect origin of personal data is from a third-party, then \textsc{PD Origin.Indirect.Third-Party} is \textit{identified}; otherwise go to \textbf{(3)}.
    \item[\textbf{(3)}] If the indirect origin of personal data is from public sources, then \textsc{PD Origin.Indirect.Publicly} is \textit{identified}; otherwise \textsc{PD Origin.Indirect} is \textit{partially identified}  -- \textbf{Warning}.
\end{itemize}

Note that  C16 in Fig.~\ref{fig:Warning} does not refer to \textsc{Cookie} although \textsc{Cookie} is a subtype of \textsc{PD Origin.Indirect} in the conceptual model of Fig.~\ref{fig:metadata}. The above-shown criterion strictly follows GDPR, which does not regulate cookies. However, our collaborating legal experts suggested the inclusion of \textsc{Cookie} in our conceptual model since cookies are often mentioned in privacy policies and they may become relevant to GDPR in the future.

\begin{figure*}[ht!]
\centering
\includegraphics[width=\linewidth] {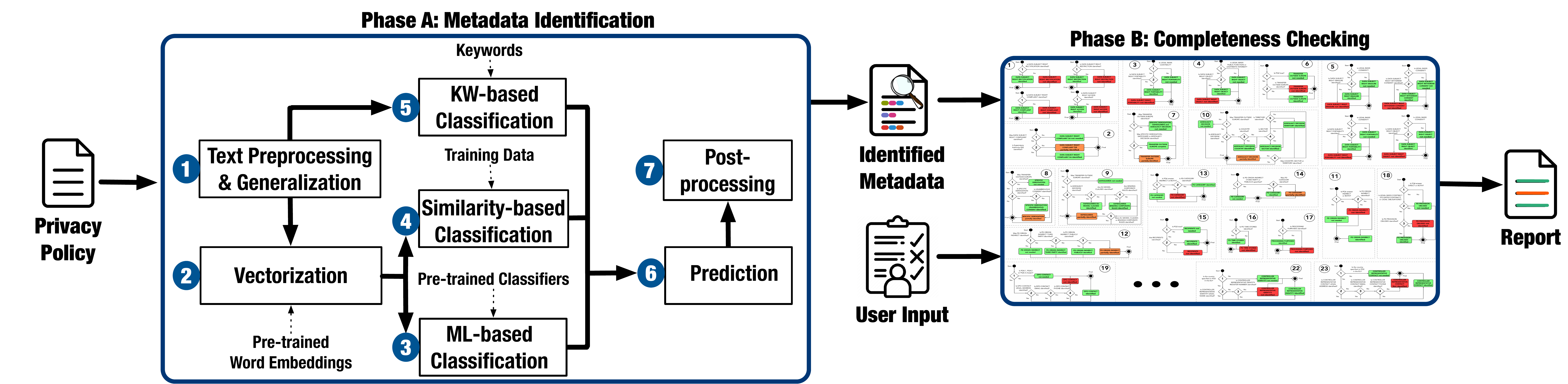}
\caption{Overview of the Completeness Checking Approach (CompAI).} \label{fig:approach}
\end{figure*}
\section{AI-based Approach for Completeness Checking (RQ3)} \label{sec:approach}

In this section, we address RQ3 and present our AI-based approach for \textbf{Comp}liance checking of privacy policies using \textbf{A}rtificial \textbf{I}ntelligence against GDPR (thereafter referred to as \textit{CompAI}). 
\textcolor{black}{ \textit{CompAI} does not use deep learning (DL) architectures (e.g., LSTM~\cite{hochreiter:97}), since we do not have enough data for developing such models with enough accuracy. We leave investigating the possibility of using DL for future work.}
CompAI, shown in Fig.~\ref{fig:approach}, is composed of two main phases.  
Phase~A, \textit{metadata identification}, takes as an input a privacy policy, and returns the metadata types that are present in this policy as an intermediary output. 
\textcolor{black}{More precisely, Phase A results in a binary decision for each metadata type regarding whether or not it is present in the input privacy policy.  }
Phase~B, \textit{completeness checking}, takes as an input the identified metadata types from Phase~A and the user input based on the questionnaire (explained in Sec.~\ref{sec:criteria}). Phase~B then returns a detailed report about whether the input privacy policy is complete according to GDPR.  
We elaborate these  phases next.

\subsection{Metadata Identification (Phase~A)} 
Phase~A uses a combination of NLP and ML to identify the metadata types that are present in a given privacy policy. 
Our metadata identification approach aims to solve a hierarchical, multi-label and multi-class classification problem. The nature of the problem is visible from the conceptual model in Fig.~\ref{fig:metadata}, where most level-1 metadata types are further specialized into sub-metadata types (level-2 and \hbox{level-3}). Multi-label classification reflects the fact that a sentence in the privacy policy can discuss one or more metadata types. Therefore, our solution can predict one or more potential labels (metadata types) for each sentence in the input privacy policy.
Our approach considers a \textit{sentence} as the unit of analysis. 
A sentence refers to the textual entity that results from applying the sentence splitting module in the NLP pipeline (Fig.~\ref{fig:pipeline}), irrespective of whether the sentence identified by this module corresponds to a grammatical sentence. 
The rationale behind using sentences rather than phrases 
is that the former are more likely to contain the context necessary for understanding their meaning~\cite{michaelis:03} and thus lead to more accurate classification results. 

Phase A is further composed of seven steps. In the first two steps, 
the text of the input privacy policy is preprocessed, generalized and transformed into a mathematical representation (vectors). 
In steps~3-5, we classify the sentences of the input privacy policy into one or more metadata types using three classification methods based on ML, semantic similarity, and keywords. \textcolor{black}{As we will see, relying on these complementary methods is necessary to overcome the complexity of the hierarchical classification problem. } 
In step~6, we combine the results of steps~3, 4, and 5 to predict metadata types for each sentence in the input privacy policy. In the last step, we refine the results through  post-processing. We explain these steps in detail next.

\subsubsection{Text Preprocessing and Generalization (Step~1)}
\begin{figure}[H]
\centering
\includegraphics[width=.95\linewidth] {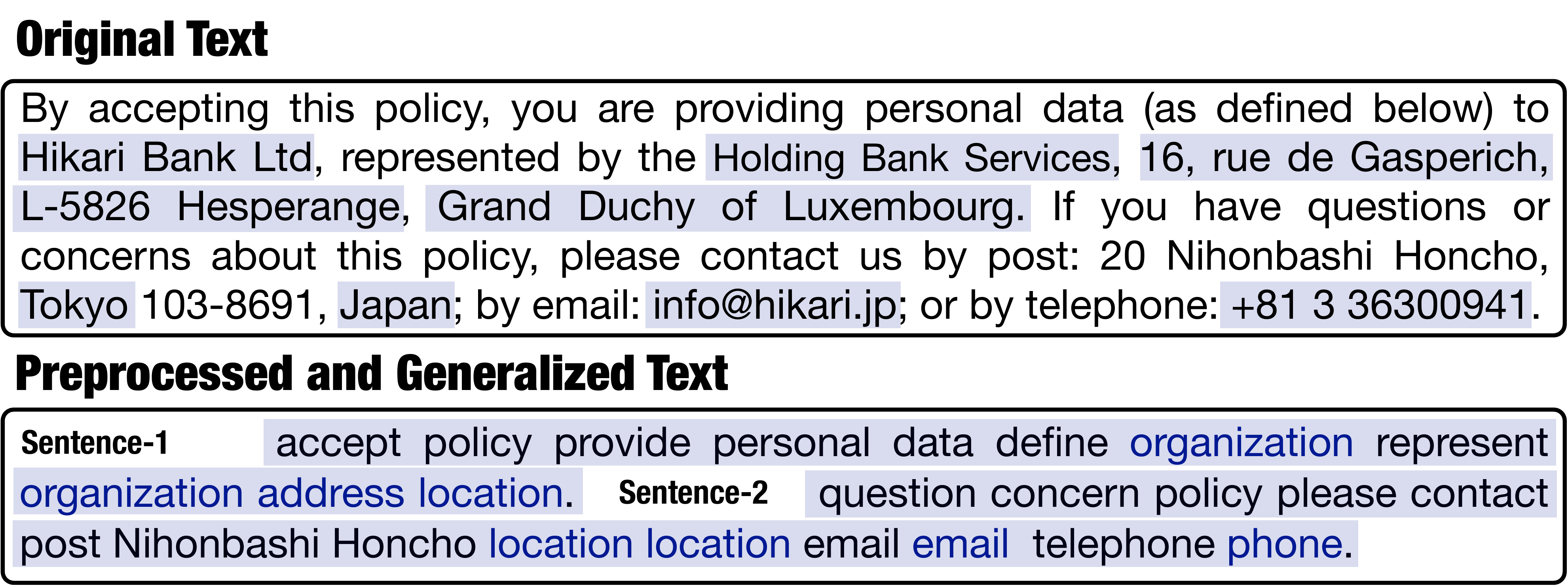}
\caption{Example of Text Preprocessing and Generalization.} \label{fig:step1}
\vspace*{-.5em}
\end{figure}

In step~1, we apply the NLP pipeline (Fig.~\ref{fig:pipeline}) to parse the input privacy policy and obtain the sentences. Using the annotations produced by the NLP pipeline, we generalize the text in each sentence by replacing specific textual entities with more generic ones. 
Specifically, we replace named entities (as identified by the named entity recognition module) 
with their types. For example, the entities ``Japan'' and ``Hikari Bank Ltd'' in Fig.~\ref{fig:step1} will be replaced with the types \textit{location} and \textit{organization}, respectively. Similarly, we generalize emails, postal addresses, telephone numbers, and websites, e.g., ``info@hikari.jp'' is replaced with \textit{email}. 
The intuition behind generalization is to normalize the text such that, despite significant diversity across the privacy policies used for training (e.g., the mention of different locations), the approach can still learn common patterns and accurately predict metadata types. 
The generalized sentences are further normalized through lemmatization and stopword removal, e.g., in  Fig.~\ref{fig:step1} ``accepting'' becomes ``accept'' and stopwords like ``by'' are removed. 

\subsubsection{Vectorization (Step~2)}
Step~2 transforms the textual sentences resulting from step~1 into embeddings. To do this, we utilize the pre-trained word-vector model of 100-dimensional vectors from 
GloVe~\cite{pennington:14} (introduced in  Sec.~\ref{subsec:embeddings}).
\textcolor{black}{Using off-the-shelf, pre-trained and context-independent (i.e., one vector per word regardless of context) word vectors increases the applicability of our approach by making it directly applicable for analyzing new document types.}
For computing the sentence embedding, we 
first retrieve the corresponding embedding for each word in the sentence as given by the pre-trained model. Then, we average over all the word embeddings to get a single vector representing the sentence embedding. For example, the embedding of the sentence ``data privacy policy'' in \begin{wrapfigure}{r} {0.27\textwidth} 
\includegraphics[width=0.27\textwidth]{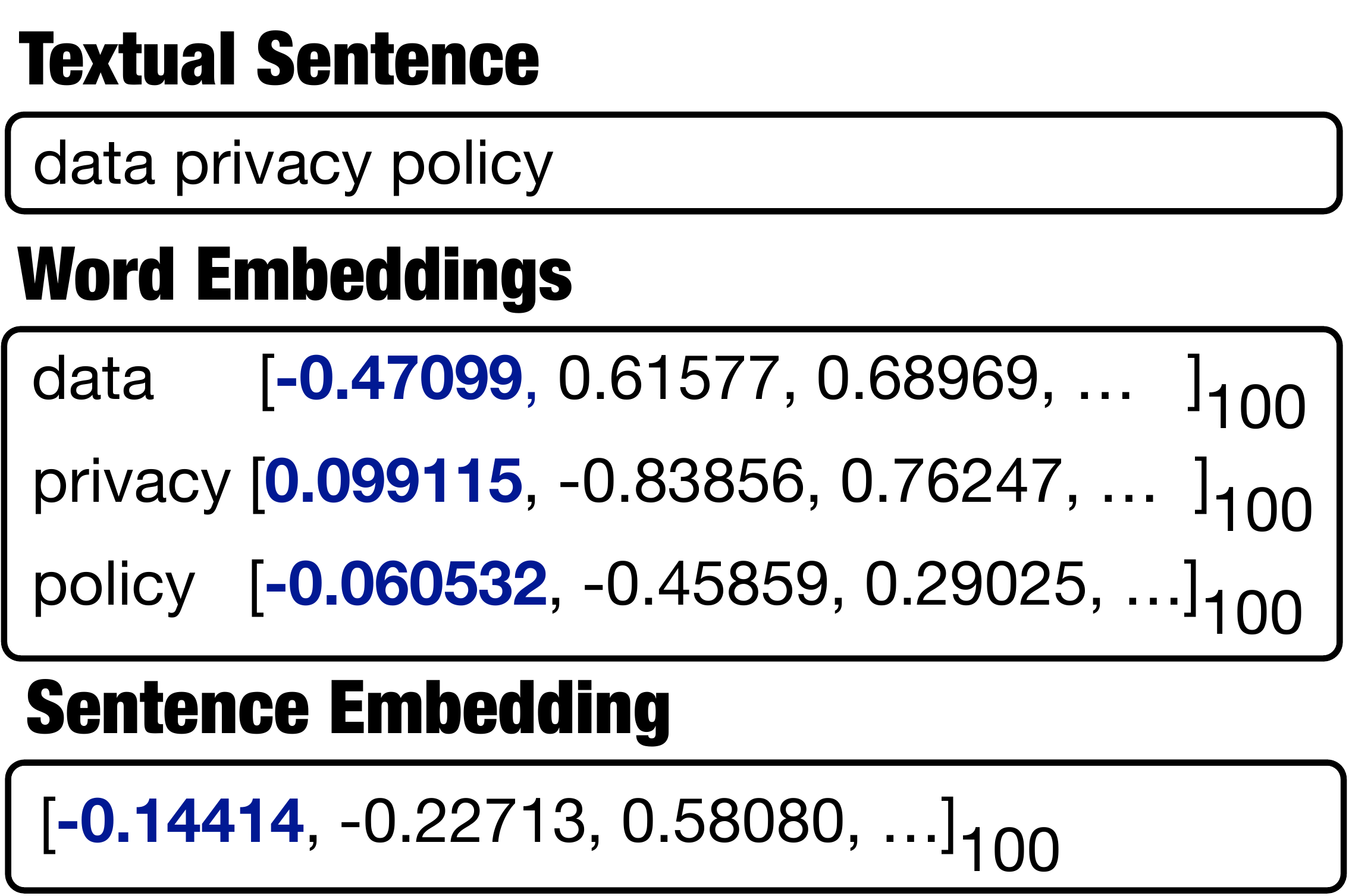}
\caption{Example of Vectorization. } \label{fig:embeddings} 
\vspace*{-.5em}
\end{wrapfigure}Fig.~\ref{fig:embeddings} is the average of the word embeddings in that sentence, such that the first entry in the sentence embedding (i.e., -0.14414) corresponds to the average of the first entries in the word embeddings of ``data'', ``privacy'', and ``policy'' (i.e., -0.47099, 0.099115, and -0.060532), respectively. 
\textcolor{black}{
The objective of the vectorization step is to achieve a representation for measuring text similarity that is effective and fast to train and test. Driven by this objective, we use simple averaging of embeddings because doing so has proven to be efficient for generating sentence embeddings across a broad range of different domains and NLP tasks, including text similarity~\cite{Wieting:15, Blacoe:12}.  }
\subsubsection{ML-based Classification (Step~3)} 
In this step, we attempt to solve the multi-class, multi-label classification problem by transforming it into multiple binary classification problems (as explained in Sec.~\ref{subsec:ml}). 
To do so, we apply the pre-trained ML classifiers for predicting the presence of level-1 and level-2 metadata types in each sentence of the input privacy policy. 
We restrict the use of ML to level-1 and level-2 metadata types 
because the number of positive examples we have in our training set for level-3 metadata types is not sufficient for building accurate ML classifiers at that level. 

Our classifiers are trained on a feature matrix in which each row corresponds to a sentence and the columns are the 100-dimensional sentence embedding computed in step~2. The prediction class for each classifier indicates the presence of a level-1 or level-2 metadata type in the sentence. For example, the sentence: \textit{Your personal data might be disclosed to the tax authorities, or other third parties including legal or financial advisors, regulatory bodies, auditors and technology providers.} (number \textbf{14} in Fig.~\ref{fig:annotations}) is predicted as 
\textsc{Recipients}. 
We train the classifiers with positive examples representing the sentences that have been annotated with a particular metadata type (e.g., \textsc{Data Subject Right}) and negative examples annotated with any other metadata type \textit{at the same level} (i.e., all but \textsc{Data Subject Right}). In most of the cases, we obtained imbalanced datasets with positive examples being under-represented.
Inspired by Wang and Manning~\cite{wang:12}, we use a support-vector machine (SVM) classifier with its default hyper-parameters for sentence classification. SVM is widely used for text classification~\cite{komninos:16}.
We address the imbalance problem in our work using under-sampling over negative examples~\cite{Witten:16}. \textcolor{black}{Our preliminary experiments suggested that using both SVM for text classification and under-sampling for handling imbalanced datasets outperformed alternatives, e.g., using Na\"ive Bayes classifier or minority over-sampling. Further, as we will discuss in Sec.~\ref{sec:evaluation}, the high accuracy obtained by our current solution alleviates the need to empirically examine alternatives.}

Step~3 uses one pre-trained binary classifier for each \hbox{level-1} and level-2 metadata type in the model of Fig.~\ref{fig:metadata}. 
The classifier predicts for each sentence in the input privacy policy, using its embedding vector as features, whether it should be labelled with the metadata type on which the classifier
has been trained. For example, the sentence: \textit{the right to request erasure, restriction, portability, and to object to the processing of your personal data}  (numbers \textbf{29} -- \textbf{32} in Fig.~\ref{fig:annotations}) is predicted by the corresponding binary classifiers as the level-2 metadata types \textsc{Data Subject Right.\{Erasure, Restriction, Portability, and Object\}}. If a metadata type is not predicted by any binary classifier to be present in a sentence, then this metadata type is deemed as absent. 
The resulting classifications are passed on to step~6 (Prediction).

\subsubsection{Similarity-based Classification (Step~4)}
In this step, we classify each sentence of the input privacy policy based on how similar it is to the group of sentences, in the training set, that are annotated with a certain level-1 or level-2 metadata type. Restricting the use of this classification to level-1 and level-2 metadata types is due to the same reason  explained in step~3. 

Step~4 creates one group for each level-1 and level-2 metadata type. Similar to step~3, this step characterizes a sentence using the vector representation built in step~2. Since an individual sentence can have multiple metadata-type annotations, the same sentence embedding can be part of several groups.
Each group is represented by a single vector which is computed as the average of all sentence embeddings in that group. To predict whether a sentence ($S$) should be annotated with a certain metadata type ($t$), we compute the \emph{cosine similarity} between the sentence embedding ($\vec{S}$) and the vector capturing the average embedding of the group of sentences annotated by $t$ (i.e., $\vec{t}$) in the training set. If the cosine similarity is above a pre-specified threshold, we predict $t$ to be a metadata type for $S$. We set the value of this threshold to 0.9. 
This value was empirically obtained by evaluating the accuracy of the prediction using a range of similarity threshold values between 0.5 and 0.9, with a step of 0.01, on a subset of the privacy policies in the training set. Threshold values less than 0.5 are not considered because they fail to capture  similarity. 

To illustrate, consider our example in Fig.~\ref{fig:annotations}. The cosine similarity between the group of sentences annotated with \textsc{PD Origin.Indirect} ($t$) and the vector representation ($\vec{S}$) of the sentence: \textit{information obtained from third parties including [...]} (number \textbf{10}) is 0.91, while the cosine similarity with $\vec{S^\prime}$ of the sentence: \textit{Your personal data might be disclosed to the tax authorities, or other third parties [...]} (number \textbf{14}) is  0.43. As a result, $S$ is classified as $t$ while $S^\prime$ is not. The results of this step are passed on to step~6 (Prediction).

\subsubsection{Keyword-based Classification (Step~5)}
In this step, we conduct a keyword search (from a predefined list) over the (textual) sentences in the input privacy policy. If a sentence $S$ contains one or more of the keywords associated with metadata type $t$, then we predict that $S$ should be annotated with $t$. 
\textcolor{black}{For example, the sentence (number \textbf{16} in Fig.~\ref{fig:annotations}): \textit{We may also transfer your personal data to countries outside the European Union (including Japan) on the basis of: European Commission's adequacy decisions, certified by the \textbf{APPI Japan Scheme}.} will be predicted as \textsc{Transfer Outside Europe.Adequacy Decision.Country}, since it contains keywords indicating this metadata type (highlighted in bold).} \textcolor{black}{Another important point in relation to keywords is that, the text generalization performed in step~1 improves the efficacy of keyword search. 
For example, number \textbf{5} in Fig.~\ref{fig:annotations} represents an email address that is generalized with \textit{email}. Thus, including \textit{email} as a keyword enables identifying the metadata type \textsc{Contact.Email}. }
\textcolor{black}{We have collected a list of keywords covering all of the metadata types in Fig.~\ref{fig:metadata}. We elaborate in Sec.~\ref{sec:evaluation} on how we obtain these keywords. } 
The results of this step are passed on to the next step (Prediction).

\subsubsection{Prediction (Step~6)}
This step combines the classification results produced based on ML (step~3), semantic similarity (step~4) and keyword search (step~5) to produce a final recommendation about which metadata types should be ascribed to a given sentence. 

The reason why we use three different classifiers is to overcome the complexity of the hierarchical multi-class classification problem and hence improve the accuracy of predicting the potential labels for each sentence in the privacy policy. 
Each method alone has some limitations. On the one hand, relying only on keyword search is not sufficient because of the limitations discussed in Sec.~\ref{sec:introduction}. ML-based and similarity-based classifications, on the other hand, are restricted to \hbox{level-1} and \hbox{level-2} metadata types and are further more accurate for the former since the number of datapoints gets much smaller at level-2. 
Thus, ensembling the three classifiers 
yields accurate predictions as we will show in our empirical evaluation (Sec.~\ref{sec:evaluation}). 

Our strategy for combining the above classification methods is elaborated in Algorithm~\ref{alg}. The algorithm applies ML-based and similarity-based classifiers for predicting both level-1 and level-2 metadata types. Despite having keywords for all metadata types, the use of keyword search in our approach is limited. We use keywords to predict level-3 that is specializing an already-predicted (level-2) metadata type or to provide supporting evidence for predicting a level-2 metadata type in case its level-1 cannot be predicted. 

The algorithm starts with an initially empty set of labels  
($\mathcal{M}$) -- \hbox{line 1}. A label can be represented as \textit{\hbox{level-1}.level-2.level-3} for specialized metadata, e.g., \textsc{Data Subject Right.Complaint.SA}. A partial label can also be predicted, e.g., \textsc{Data Subject Right.Complaint} or  \textsc{Children}, in case the metadata has no specialization or there is no evidence that supports predicting a specialization. 

\renewcommand{\algorithmiccomment}[1]{\bgroup\hfill//~#1\egroup}
\setlength{\textfloatsep}{3pt}
\begin{algorithm}[t!]
\caption{Metadata Prediction for a Sentence $S$}\label{alg}
\begin{algorithmic}[1]
\REQUIRE  $\vec{S}$:~vector representation of $S$; 
$\mathit{cf}_1$,~$\mathit{cf}_2$:~binary classifiers trained on level-1 and level-2 metadata types for $\vec{S}$, respectively; 
$\vec{av}({t})$:~average vector for the group of sentences annotated with metadata type $t$; 
$\mathcal{K}$:~set of metadata types predicted based on keyword search in $S$;  
$C_{ID}$, $CR_{ID}$:~the values of \textsc{Controller.Identity} and \textsc{Controller Representative.Identity}, respectively.
\ENSURE $\mathcal{M}$:~a set of metadata types predicted for $S$

\STATE $\mathcal{M} \leftarrow \emptyset$ 
\STATE Let $\mathcal{L}_1$ be the set of level-1 metadata types 
\FOR{$\ell_i \in \mathcal{L}_1$}
    \IF[\textcolor{blue}{\it \hbox{Predict level-1 \& level-2 (Case~1)}}]{$\mathit{cf}_1$ predicts $\ell_i$ \OR $\mathtt{sim}(\vec{S}, \vec{av}(\ell_i))\geq0.9$} 
        \STATE Add $\ell_i$ to $\mathcal{M}$
        \FOR{$\ell_j$ s.\,t. $\ell_j$ is a (level-2) specialization of $\ell_i$}
            \IF{$\mathit{cf}_2$ predicts $\ell_j$ \OR $\mathtt{sim}(\vec{S}, \vec{av}(\ell_j))\geq0.9$}
                \STATE Add $\ell_i.\ell_j$ to $\mathcal{M}$
            \ENDIF
        \ENDFOR
    \ELSE[\textcolor{blue}{\it Predict level-1 \& level-2 (Case~2)}]
        \FOR{$\ell_j$ s.\,t. $\ell_j$ is a (level-2) specialization of $\ell_i$}
            \IF{($\mathit{cf}_2$ predicts $\ell_j$ \AND $\mathtt{sim}(\vec{S}, \vec{av}(\ell_j))\geq0.9$) \OR
            \\ ($\mathit{cf}_2$ predicts $\ell_j$ \AND $\ell_j \in \mathcal{K}$)
            \OR \\($\mathtt{sim}(\vec{S}, \vec{av}(\ell_j))\geq0.9$ \AND $\ell_j \in \mathcal{K}$)
            }
                \STATE Add $\ell_i.\ell_j$ to $\mathcal{M}$
            \ENDIF
        \ENDFOR
    \ENDIF
\ENDFOR
\IF{$S$ contains $C_{ID}$}
\STATE Add \textsc{Controller.Identity} to $\mathcal{M}$
\ELSIF{$S$ contains $CR_{ID}$}
\STATE Add \textsc{Controller Representative.Identity} to $\mathcal{M}$
\ENDIF

\FOR[\textcolor{blue}{\it Predict level-3}]{$\ell_i.\ell_j \in \mathcal{M}$}
    \FOR{$\ell_q$ s.\,t. $\ell_q$ is a (level-3) specialization of $\ell_j$} 
        \IF {$\ell_q \in \mathcal{K}$}
            \STATE Add $\ell_i.\ell_j.\ell_q$ to $\mathcal{M}$
        \ENDIF  
    \ENDFOR
\ENDFOR
\end{algorithmic}
\end{algorithm}

\sectopic{Level-1 and Level-2 Metadata.} 
The algorithm predicts a level-1 metadata type and its corresponding level-2 specializations in two cases, Case~1 (lines 4 -- 10) and Case~2 (lines 11 -- 17). Case~1 applies when some \hbox{level-1} metadata type can be predicted; the algorithm then attempts to predict its level-2 type. 
Case~2 applies when Case~1 fails to predict a level-1 type but there is strong support for predicting its level-2 type. \textcolor{black}{The rationale behind Case~2 is that when two classifiers jointly predict a level-2 metadata type (as we elaborate next), then their predictions should compensate for the absence of a level-1 prediction in Case~1. If Case~2 leads to a prediction of a certain level-2 metadata type (e.g., \textsc{Vital Interest}), then this will be considered as an indirect indication for predicting the level-1 of that metadata type (e.g., \textsc{Legal Basis}). }

\emph{\textbf{Case 1:}}
\textcolor{black}{If a level-1 metadata type ($\ell_i$) is predicted for the sentence ($S$) via the (level-1) ML-based classifier ($\mathit{cf}_1$) or by the similarity-based classifier (line 4), 
then $\ell_i$ is added to $\mathcal{M}$ (Line 5). 
If the predicted $\ell_i$ has any specialization, the algorithm attempts to further predict its level-2 metadata type ($\ell_j$). 
If $\ell_j$ is predicted by the (\hbox{level-2}) ML-based ($\mathit{cf}_2$) or similarity-based classifiers (line 7), then the annotation $\ell_i.\ell_j$ is added to $\mathcal{M}$. Since $\ell_i$ has been confirmed earlier, it is sufficient to get $\ell_j$ predicted by one classifier (excluding keyword-based for the reasons mentioned earlier). Regardless of whether or not the algorithm succeeds to predict $\ell_j$, $\ell_i$ is still added to $\mathcal{M}$ (line 5). The rationale is that pinpointing the sentence that discusses $\ell_i$ helps the legal experts easily locate $\ell_j$ which is expected to appear in the following sentences.
}


\emph{\textbf{Case 2: }}
\textcolor{black}{If the level-1 metadata type ($\ell_i$) cannot be directly predicted, 
the algorithm checks whether its \hbox{level-2} ($\ell_j$) can still be predicted. Case~2 requires $\ell_j$ to be predicted by two classifiers (line 18). Specifically, the label $\ell_i.\ell_j$ is added to $\mathcal{M}$ if at least one of the following three pre-conditions is satisfied:  $\ell_j$ is predicted by the (\hbox{level-2}) ML-based ($\mathit{cf}_2$) and similarity-based classifiers (line 13 -- first condition). Alternatively,  $\ell_j$ is predicted by either $\mathit{cf}_2$ or the similarity-based classifier, and $\ell_j$ is further predicted by keyword search (line 13 -- second two conditions). In Case~2, $\ell_i$ is automatically added to the set of annotations to get the hierarchical label, since there is enough evidence to support the prediction of $\ell_j$.} 
\textcolor{black}{To obtain a joint prediction by the three classifiers in Case~2, we considered all possible combinations as described in the set of rules -- line 13. These rules include combining the predictions of (i) ML-based with similarity-based, (ii) ML-based with keyword-based, and (iii) similarity-based with keyword-based. }

The \hbox{level-2} metadata types \textsc{Controller.Identity} ($C_{ID}$) and \textsc{Controller Representative.Identity} ($CR_{ID}$) are provided by the user through the questionnaire explained in Sec.~\ref{sec:criteria} (as answers to Q1 and Q5). If
$C_{ID}$ (or $CR_{ID}$) occurs in the sentence $S$, then \textsc{Controller.Identity} (or \textsc{Controller Representative.Identity}) is added to $\mathcal{M}$ \hbox{(lines 19 -- 23)}.

\sectopic{Level-3 Metadata.} Recall that ML-based and similarity-based classifiers are not applicable to level-3 metadata types due to the lack of positive examples in our training data. Therefore, we use keyword-based classification only. 
The algorithm attempts to predict \hbox{level-3} metadata types based on any already predicted level-1.level-2 annotation. Specifically, the algorithm considers all level-3 metadata types that specialize some level-2 metadata type already in $\mathcal{M}$. 
For each level-2 metadata type that is predicted, if its level-3 
is predicted by keyword search, then level-3 metadata type is added to $\mathcal{M}$ (line 27). 

\subsubsection{Post-processing (Step~7)}
In the seventh and final step of our metadata identification approach, we refine the results of step~6 by considering the metadata types predicted for the sentences surrounding a given sentence. 
The intuition behind this step is the observation that specializations of certain metadata types are discussed in consecutive sentences of privacy policies.
Based on this observation, when a sentence $S$ is predicted as having a specific metadata type $t$, the surrounding context, specifically the preceding and succeeding sentences, can provide a confirmatory measure about whether $t$ is a reliable prediction for $S$.

We employ several such context-based heuristics for post-processing \textcolor{black}{ \textsc{Data Subject Right}, \textsc{Transfer Outside Europe}, and \textsc{Legal Basis}, since these types are often discussed in consecutive sentences in the privacy policy. }
\textcolor{black}{The heuristic states that if some level-2 metadata type ($\ell_j$) 
is predicted for a sentence ($S$),  
then we look at the $n$ 
preceding and $n$  
succeeding sentences, such that $n$ equals the number of the metadata types at the same level of $\ell_j$. \color{black} The number $n$ accounts for the possibility to discuss the level-2 of a metadata type each in a separate sentence.  
} 
\textcolor{black}{For example, eight sentences before and after a sentence are considered to belong to the context for the level-2 metadata type \textsc{Data Subject Right.Portability}, \color{black} where the \hbox{level-2} metadata types of \textsc{Data Subject Right} can be listed in eight sentences at most. 
}

If none of these surrounding sentences are predicted to discuss a metadata type relevant to $\ell_j$, 
then we remove from the annotations for $S$ the predicted label that includes $\ell_j$.  
This is because the context around $S$ lends no support to $\ell_j$ 
being a correct annotation for~$S$. \textcolor{black}{A metadata $\ell_j'$ is said to be relevant to $\ell_j$ if it belongs to the same level-1 metadata type. }
\textcolor{black}{To illustrate, let $S$ be sentence number \textbf{16} in Fig.~\ref{fig:annotations}. This sentence can be falsely classified as \textsc{Data Subject Right.Portability}, because of the misleading words (``transfer'', ``personal'', ``data'').
In post-processing, we look at the metadata types predicted for the \textit{eight} preceding (i.e., \textbf{8 -- 15}) and \textit{eight} following sentences (i.e., \textbf{17 -- 24}) to decide if there is  enough support to confirm the prediction of $S$. If none of the predicted metadata types in the context is relevant to \textsc{Data Subject Right.Portability}, then we filter out this prediction assuming it is false. }

\subsection{Completeness Checking (Phase~B)}

\begin{figure}
\centering
\includegraphics[width=\linewidth] {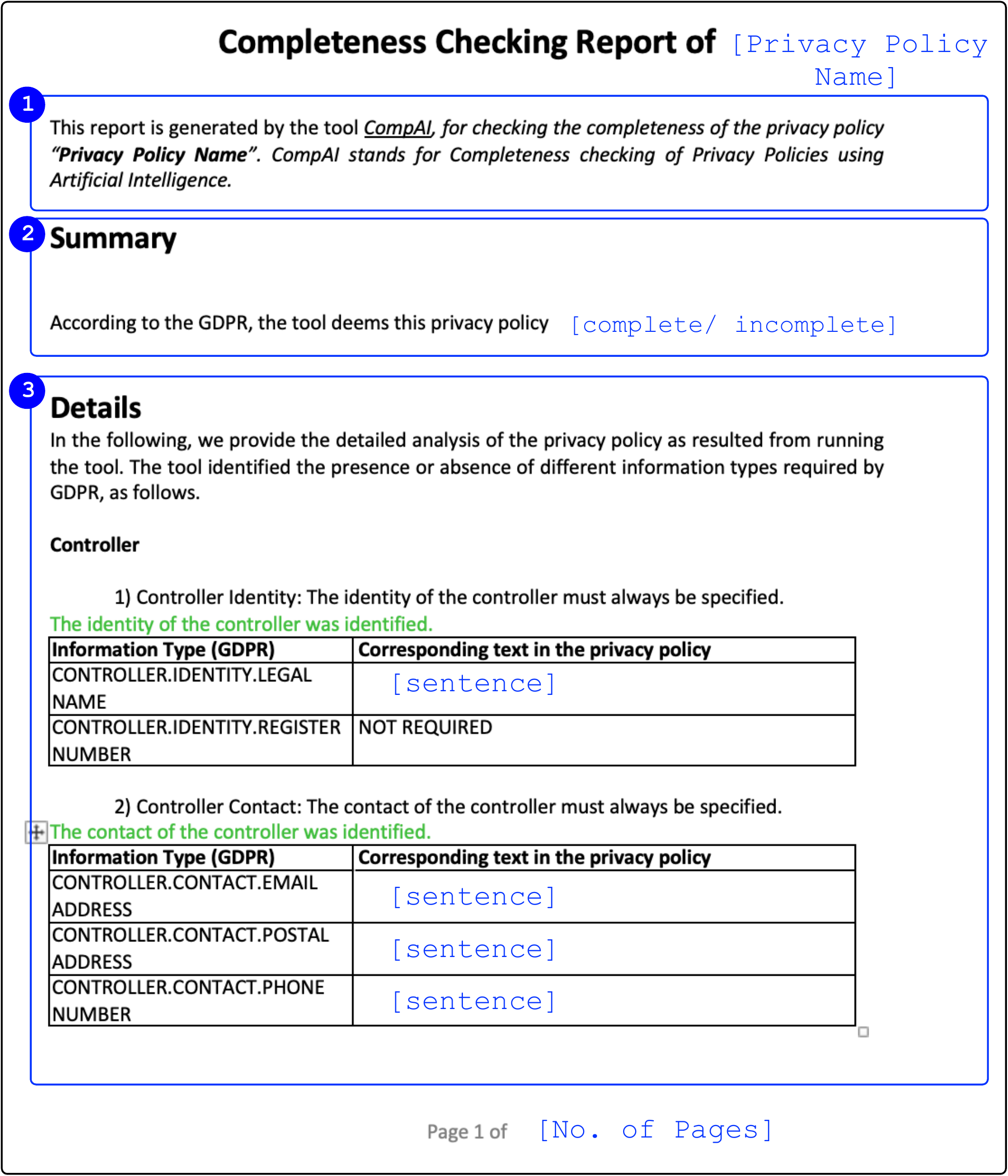}
\caption{\textcolor{black}{Template of Completeness Analysis Report.}} \label{fig:report}
\end{figure}

Phase~B takes as an input: (1) the output of Phase~A representing the predicted metadata types in the input privacy policy, and (2) the answers of the user to the six questions discussed in Sec.~\ref{sec:criteria}. Phase~B then returns a detailed report on completeness analysis as the final output of our overall approach. \textcolor{black}{Fig.~\ref{fig:report} shows the template of the report which CompAI generates. The first part is a preamble including the name of the privacy policy. The second part presents a summary about the final decision regarding completeness. The third part shows the details of the identified metadata types under each completeness criterion. If the metadata type is not identified in any sentence, the report will show ``NOT FOUND'' and indicate a violation or warning accordingly. If the metadata type is not required because the presence of another metadata type is sufficient, or if the criterion is not applicable based on the answers to the questionnaire, then the report will reflect this through the  respective statements ``NOT REQUIRED'' or ``NOT APPLICABLE''. }

This phase implements the completeness criteria shown in Fig.~\ref{fig:Violations} and Fig.~\ref{fig:Warning}. 
Using our running example in Fig.~\ref{fig:annotations}, the expected answers to the questionnaire are the following. The \textsc{Controller.Identity} is \textit{Hikari Bank Ltd} (\textbf{Q1)}, personal data will likely be transferred outside the EU (\textbf{Q2}), there will be recipients other than the \textsc{Controller} (\textbf{Q3}), the core activities include processing special categories (\textbf{Q4}), processing of personal data will take place in Europe (\textbf{Q5}), and finally the personal data will be collected both directly and indirectly (\textbf{Q6}). The answer to \textbf{Q5} requires an additional input from the user about the \textsc{Controller Representative.Identity} which is \textit{the Holding Bank Services}.

Based on the answers given above, all completeness criteria (see Sec.~\ref{sec:criteria}) need to be checked in this step. For example, the criterion \textbf{C22} states that \textsc{PD Provision Obliged} should be present in a privacy policy when the answer to Q6 is either \textsc{PD Origin.Direct} or both, and at the same time the legal basis of processing personal data is either \textsc{Legal Basis.Legal Obligation} or \textsc{Legal Basis.Contract.To Enter Contract}. A violation of this criterion 
raises an incompleteness issue. In our example in Fig.~\ref{fig:annotations}, both the above-mentioned metadata types are found in the privacy policy, in sentences \textbf{20} and \textbf{25}, respectively. As a result, we have to find 
the metadata type \textsc{PD Provision Obliged} in the same policy; this comes in sentence \textbf{26}. Had this sentence not been correctly identified by phase~A, either due to inaccurate prediction or because it is actually missing in the policy, then this criterion would have been violated.
The result of Phase~B is a 
set of detected violations and warnings for the 23 criteria due to missing metadata in the input privacy policy. 

\section{Empirical Evaluation}\label{sec:evaluation}

\subsection{Implementation}
We have implemented our approach using Java. The
implementation has $\approx$~7500 lines of code excluding comments and third-party libraries. For the basic NLP pipeline, we use the DKPro toolkit~\cite{castilho:14}. 
For text generalization, we use regular expressions available in Java. 
We transform words into embeddings by utilizing the publicly available pre-trained word embeddings from GloVe~\cite{pennington:14}. Noting that our implementation is Java-based, we perform operations on word embeddings using  Deeplearing4j~\cite{deeplearning4j}. 
Our metadata identification approach uses ML-based classification. For classification and handling imbalance in our dataset, we employ WEKA~\cite{Hayes:14,eibe:2016}. For computing  similarity between two textual entities in the similarity-based classification, we use Cosine Similarity~\cite{Manning:08}. 



\subsection{Data Collection Procedure} \label{subsec:dataCollection}
\begin{table*}[!t]
\centering
\caption{Document Collection Results.} 
\begin{tabular}{c p{0.23\linewidth} c c | c c | c c} 
&& \multicolumn{2}{c}{\textbf{Total}}&\multicolumn{2}{c}{Training Data (\textit{T})} & \multicolumn{2}{c}{Test Data (\textit{E})} \\
\midrule
\textbf{Level} & \textbf{Metadata} & \textbf{Manifestations} & \textbf{Sentences} &
\textbf{Manifestations} & \textbf{Sentences} 
& \textbf{Manifestations} & \textbf{Sentences} \\
\toprule
L1&\textbf{\textsc{Controller}}&-&-&-&-&-&-\\
L2&\textsc{Identity} & 221 & 799 & 175 & 415 & 46 & 384 \\
L2&\textsc{Contact} & 151 & 652 & 117 & 466 & 34 & 186 \\
\midrule
L1&\textbf{\textsc{Controller Representative}}&-&-&-&-&-&-\\
L2&\textsc{Identity} & 19 & 36 & 16 & 33 & 3 & 3 \\
L2&\textsc{Contact} & 21  & 63 & 18&60  & 3 &3 \\
\midrule
L1&\textbf{\textsc{DPO}} & - & - & - & - & - & -  \\
L2&\textsc{Contact} & 125 & 462 & 104 & 404 & 21 & 58 \\
\midrule
L1&\textbf{\textsc{Data Subject Right}} & 224 & 3352 & 180 & 2651 & 44 & 701\\
L2&\textsc{Access} & 209 & 378 & 167 & 304 & 42 & 74 \\
L2&\textsc{Rectification} & 211 & 345 & 169 & 267 & 42 & 78 \\
L2&\textsc{Restriction} & 170 & 311 & 137 & 247 & 33 & 64 \\
L2&\textsc{Complaint} & 172 & 286 & 137 & 219 & 35 & 67 \\
L3&\textsc{SA} & 172 & 264 & 138 & 217 & 34 & 47 \\
L2&\textsc{Erasure} & 196 & 386 & 157 & 296 & 39 & 90 \\
L2&\textsc{Object} & 181 & 484 & 145 & 373 & 36 & 111 \\
L2&\textsc{Portability} & 163 & 263 & 131 & 210 & 32 & 53 \\
L2&\textsc{Withdraw Consent} & 169 & 395 & 136 & 322 & 33 & 73\\
\midrule
L1&\textbf{\textsc{Legal Basis}} & 231 & 4511 & 185 & 3238 & 46 & 1273 \\
L2&\textsc{Contract} & 161 & 553 & 127 & 366 & 34 & 187 \\
L3&\textsc{Contractual} & 123 & 275 & 89 & 183 & 34& 92  \\
L3&\textsc{To Enter Contract} & 73 & 105 & 18 & 30 & 55 & 75 \\
L3&\textsc{Statutory} & 20 & 25 & 13 & 16 & 7 & 9\\
L2&\textsc{Public Function} & 73 & 122 & 51 & 84 & 22 & 38 \\
L2&\textsc{Legitimate Interest} & 214 & 2424 & 170 & 1846 & 44 & 578 \\
L2&\textsc{Vital Interest} & 17 & 24 & 10&	15 & 7	&9 \\
L2&\textsc{Consent} & 180& 554 & 141 & 423 & 39& 131 \\
L2&\textsc{Legal Obligation} & 200& 1028 & 155 & 704 & 45 & 324 \\
\midrule
L1&\textbf{\textsc{Transfer Outside Europe}} & 178 & 823 & 148 & 707 &30 & 116 \\
L2&\textsc{Adequacy Decision} & 47& 76 & 64 & 109 & 4 & 4 \\
L3&\textsc{Country} & 47 & 76 & 45 & 74 & 2 &2 \\
L2&\textsc{Safeguards} &136 &280 & 113 & 230 & 23 & 50 \\
L3&\textsc{Binding Corporate Rules} & 50 & 64 & 46 & 58& 4 & 6 \\
L3&\textsc{EU Model Clauses} &96& 129 & 76 & 100 & 20 & 29 \\
L2&\textsc{Specific Derogation} & 17 &20 & 10 & 13 & 7 & 7 \\
L3&\textsc{Unambiguous Consent} & 16 & 18& 10 & 12 & 6 & 6 \\
\midrule
L1&\textbf{\textsc{PD Origin}} & 216 & 1904 & 310 & 125 & 45 & 356\\
L2&\textsc{Direct} & 165 & 436 & 125 & 310 & 40 & 126 \\
L2&\textsc{Indirect} & 209 &1356 & 164 & 1129 & 45 & 227 \\
L3&\textsc{Third Party} & 113& 294 & 80 & 206 & 33 & 88  \\
L3&\textsc{Publicly} & 79 & 127& 55& 80 & 24 & 47  \\
L3&\textsc{Cookie} & 155 & 668 & 123 & 597 & 32 & 71 \\
\midrule
L1&\textbf{\textsc{PD Category}} & 228 & 2209 & 182 & 1860 & 46& 349 \\
L2&\textsc{Special} & 98 & 265& 69	& 198 & 29 & 67 \\
L2&\textsc{Type} & 33 & 71 &24&	60 & 9	&11\\
\midrule
L1&\textbf{\textsc{Recipients}} & 209& 1599& 167	&1369 & 42&	230  \\
\midrule
L1&\textbf{\textsc{PD Time Stored}} & 200 & 873 & 162	&738 & 38	&135 \\
\midrule
L1&\textbf{\textsc{PD Provision Obliged}} & 128 & 281& 102 & 230 & 26 & 51 \\
\midrule
L1&\textbf{\textsc{Processing Purposes}} & 158 & 1422 &  112&	1099 &46&	323\\
\midrule
L1&\textbf{\textsc{PD Security}} & 182 & 883 & 140&	717 & 42&	166 \\
\midrule
L1&\textbf{\textsc{Auto Decision Making}} & 84 & 295 & 63&	224 & 21&	71 \\
\midrule
L1&\textbf{\textsc{Children}} & 24 & 70 & 19	&58 & 5 &	12
\\
\bottomrule
\end{tabular}
\label{tab:docCollection}
\vspace*{-1.7em}
\end{table*}
Our data collection aimed at collecting and annotating privacy policies according to the conceptual model of Fig.~\ref{fig:metadata}. 
Specifically, we collected from the fund domain a total of 234 privacy policies, of which about 60\% were provided to us by Linklaters. For the remaining 40\%, we downloaded privacy policies from companies in the fund registry of Luxembourg, which has a substantial footprint in fund management~\cite{ALFI19}. 
We chose the fund domain because it is one of the main domains in which Linklaters 
is active. 
Focusing on the fund domain has an impact on the external threat to validity, as we will elaborate in Sec.~\ref{sec:threats}. 
Nonetheless, the conceptual model described in Sec.~\ref{sec:qualitative} is domain-agnostic, noting that it was derived from GDPR and the (domain-independent) knowledge of legal experts about privacy policies. 

Our data collection was performed in two steps. In the first step, a batch of 30 
policies was annotated by \textcolor{black}{the third  author} of this paper who has acquired domain expertise through close interaction with Linklaters. For annotating this first batch, hypothesis coding was applied (as explained in Sec.~\ref{sec:qualitative}). During this step, we also drafted detailed guidelines \textcolor{black}{with illustrative examples to explain the annotation process. These guidelines were then shared with the external annotators}.

The second batch \textcolor{black}{(204 policies)} was annotated by four third-party annotators (non-authors). 
Three of these individuals are graduate students in social sciences; they are native English speakers with considerable prior exposure to legal documents. The fourth annotator is a computer-science graduate student with an excellent command of English and six months of prior internship experience \textcolor{black}{on legal text processing in industry.} All four annotators attended two four-hour training sessions, focused on GDPR concepts and the definitions of our metadata types. The annotators were further provided with the guidelines drafted in the first step, 
\textcolor{black}{ during which the conceptual model was also refined. To obtain an unbiased evaluation, the conceptual model was frozen before the second step started since a subset of the annotations is used for evaluating our approach as we explain later in this section. Thus, the two steps of our annotation process are performed in a strict sequence.}
During the entire annotation process, the annotators kept track of the keywords that were frequently used to express certain metadata types. 

The annotators were asked to annotate each sentence in the privacy policies with the metadata types that they deemed to be present in the sentence. When no metadata was present, \textcolor{black}{they classified the sentence as \textit{no metadata identified}.}
To illustrate, consider the example in Fig.~\ref{fig:annotations}. Numbers \textbf{27} -- \textbf{34} represent one sentence \textcolor{black}{that includes} multiple metadata types related to \textsc{\textcolor{black}{Data Subject Right}}. The annotators would then annotate the sentence with \emph{all} the metadata types that are present in the sentence. 
%
\textcolor{black}{To measure the quality of our dataset, we computed the interrater agreement using Cohen's Kappa ($\kappa$)~\cite{Cohen:60}. Specifically, we selected 24 privacy policies ($\approx$10\% of our dataset) using random stratification to ensure that this subset covers some annotations from each of the four third-party annotators. The annotated sentences in the 24 privacy policies were independently checked by the first author who had done more than half a year of training on the completeness checking of privacy policies before validating the annotations.
The interrater agreement is computed for level-1 metadata types only. 
The agreement obtained at a sentence-level is on average $\kappa=0.71$ indicating ``moderate agreement''~\cite{Mchugh:12}. 
We observed that most of the disagreements occurred over the identification of \textsc{Processing Purposes} and \textsc{Legal Basis}. Since these two metadata types are usually related and often span multiple sentences in a privacy policy, such disagreements are expected. \textsc{Legal Basis} ensures that the processing of personal data for certain purposes (i.e., \textsc{Processing Purposes}) is lawful when some conditions are satisfied in line with the sub-metadata of \textsc{Legal Basis}. 
At a privacy-policy level, we obtained an average $\kappa$ score of 0.87, indicating ``strong agreement''~\cite{Mchugh:12}. This suggests that the annotators strongly agreed on which metadata types are present in a given privacy policy. We believe this agreement is acceptable in our context given that our automated solution aims at identifying metadata at a privacy-policy level. } 

Table~\ref{tab:docCollection} shows the results of our document collection. 
Following best practices, the entire document collection (234 privacy policies) is split randomly into two subsets containing about 80\% and 20\% of the policies, respectively used for training and development (186 policies) and for evaluation (48 policies). The first batch used in our annotation for finalizing the model is included in the training set. Hereafter, we refer to the dataset used for training as \textit{T}, and to that used for testing as \textit{E}. We use \textit{E} for answering the research questions (RQs).

The table provides statistics about the entire dataset, \textit{T} and \textit{E}. 
\textcolor{black}{Specifically, we provide per metadata type $t$: the number of manifestations of $t$ in our document collection, and the number of sentences that are annotated with $t$. A \textit{manifestation} of $t$ is counted once per privacy policy when $t$ appears in that privacy policy (i.e., there is at least one sentence annotated with $t$). We compute the number of manifestations of $t$ in our document collection as the sum of the manifestations of $t$ across the privacy policies. 
For example, the number of manifestations of \textsc{\textcolor{black}{Data Subject Right}} in Table~\ref{tab:docCollection} is 224 (i.e., \textsc{\textcolor{black}{Data Subject Right}} appears in 224 out of 234 privacy policies). Further, this metadata type was annotated in a total of 3352 sentences across the privacy policies in our collection. }
We note that none of the sentences in our dataset is annotated with \textsc{Controller}, \textsc{Controller Representative} or \textsc{DPO} as separate labels. These metadata types 
always appear 
with their specializations, e.g., \textsc{Controller.Identity} or \textsc{Controller.Contact}.


\subsection{Evaluation Procedure} \label{subsec:evalProcedure}

We answer RQ4 -- RQ6 by conducting the experiments explained next. 

\sectopic{EXPI.} This experiment answers RQ4. We assess the accuracy of our metadata identification approach. 
To do so, we run our approach and compare the results against manual annotations of the privacy policies in the test set \textit{E} 
(defined in Sec.~\ref{subsec:dataCollection}). 
We evaluate, in EXPI, 
\textcolor{black}{the manifestations of metadata types detected 
by our approach 
in a given privacy policy.} 
\textcolor{black}{We recall that a manifestation of a metadata type is counted only once per privacy policy, even if the metadata type appears in multiple sentences. } 
%
Designing our evaluation around \textcolor{black}{manifestations} 
(instead of actual sentences) is driven by our objective, which is completeness checking. To verify the completeness criteria, presented in Sec.~\ref{sec:criteria}, one needs to ascertain whether or not 
\textcolor{black}{a manifestation of the metadata type exists 
in the privacy policy.} 
To illustrate, consider criterion \textbf{C6} as an example. In C6, we need to find \textsc{Data Subject Right.Portability} in the privacy policy, when the legal ground is based on \textsc{Contract}. If our approach is able to identify \textcolor{black}{manifestations} 
of \textsc{Contract} and \textsc{Portability}, then the completeness of the policy can be properly checked. 

\textcolor{black}{Let 
a manifestation 
of the metadata type $t_i$ be represented, in a privacy policy, by $\mathcal{S}= \{s_1, s_2, \ldots, s_n\}$, such that $\mathcal{S}$ is the set of sentences that are annotated with $t_i$ according to our ground truth. 
\textcolor{black}{Our approach deems a manifestation of $t_i$ as present, if $t_i$ is predicted as a label for at least one sentence in the privacy policy. }
Following this, we define a \textit{True Positive (TP)} when the approach correctly identifies a manifestation 
of $t_i$, i.e., the approach finds at least one sentence $s_j \in \mathcal{S}$. A \textit{False Positive (FP)} is when the approach falsely identifies a manifestation 
of $t_i$, i.e., the approach finds a group of sentences $\mathcal{S}'$ to be about $t_i$, such that either $\mathcal{S}' \nsubseteq \mathcal{S}$ or 
there is no manifestation of $t_i$ 
in the privacy policy according to our ground truth. 
A \textit{False Negative (FN)} is when the approach misses a manifestation 
of $t_i$, i.e., the approach does not find any $s_j \in \mathcal{S}$. }

In EXPI, we report the overall
\textit{Accuracy (A)}, \textit{Precision (P)}, \textit{Recall (R)}, and the harmonic mean \textit{F-measure (F$_\beta$)} for each metadata type across \textit{E}. 
We compute these metrics as $A=(\mathit{TP} + \mathit{FN}) /(\mathit{TP} + \mathit{FP} + \mathit{FN} + \mathit{TN})$, $P=\mathit{TP} /(\mathit{TP} + \mathit{FP})$, $R=\mathit{TP} /(\mathit{TP} + \mathit{FN})$, and $F_\beta=(1+\beta^2)*(P*R)/(\beta^2*P+R)$.
For metadata identification, recall is more important than precision, since the metadata types identified by the approach will be used to check completeness of privacy policies. This means that, if a metadata type is falsely introduced by the approach, it can be reviewed and filtered out by an analyst, whereas missing metadata types will require the analyst to review the entire privacy policy. In EXPI, we report the \textit{F2-measure} (i.e., $\beta$ = 2) to show the evaluation in favor of recall. We choose F2-measure for two reasons: First, values of $\beta \geq 2$ do not change the reasoning about our evaluation. Second,  despite recall being more important, precision still has a great value, a very low precision (too many false positive errors) will require more time and effort in filtering the erroneous findings.  

\sectopic{EXPII.} This experiment answers RQ5. We evaluate the accuracy of checking the completeness criteria on our test set. 
The unit of evaluation in EXPII is \textit{an incompleteness issue} resulting from an unsatisfied criterion in a given privacy policy. Correspondingly, we redefine a \textit{TP} as an incompleteness issue found correctly by our approach, a \textit{FP} as an incompleteness issue found by our approach when the criterion is satisfied, a \textit{FN} as an incompleteness issue missed by our approach, and a \textit{TN} when our approach correctly concludes that there is no incompleteness issue in the privacy policy. Similar to EXPI, we report \textit{A}, \textit{P}, \textit{R}, and \textit{F2-measure}. 

\sectopic{EXPIII.} This experiment answers RQ6. We compare our \textit{AI-based} approach for completeness checking to a simple approach that uses keyword search (hereon, referred to as \textit{KW-based}). 
\textcolor{black}{The latter predicts a manifestation of 
a certain metadata type $t_i$, in a given privacy policy,} if at least one keyword associated with $t_i$ is present in any sentence in this policy. We note that keyword search is introduced as one of the classifiers in our AI-based approach (Step~5 in Fig.~\ref{fig:approach}). To have a fair comparison, the list of keywords used in our approach is the same one used in the baseline. 
In EXPIII, we compare our approach against  KW-based using the same evaluation metrics defined in EXPI for metadata identification, and in EXPII for completeness checking.

\subsection{Results and Discussion} \label{subsec:RQs}
In this section, we describe the results and answer the RQs stated in Sec.~\ref{sec:introduction}.
\sectopic{\textcolor{black}{RQ4. How accurate is our proposed approach in identifying GDPR-relevant metadata in privacy policies?}}

Table~\ref{tab:resultsOfMetadaIdentification}, on the left-hand side,  shows the results of EXPI. As explained in Sec.~\ref{subsec:evalProcedure}, the results are obtained by running our metadata identification approach on the test set (\textit{E}) which is comprised of 48 privacy policies. The table reports the accuracy, precision, recall and F2-measure computed \textcolor{black}{on 
manifestations of the metadata.} We show in Table~\ref{tab:docCollection} the total number of \textcolor{black}{manifestations } 
of each metadata 
in \textit{E}.
Out of the 56 metadata types (see Fig.~\ref{fig:metadata}), we exclude the evaluation of \textsc{Controller.Identity} and \textsc{Controller Representative.Identity} because they are given as input by the user, and are looked up in the privacy policy  rather than being identified like the other metadata types (see Algorithm~\ref{alg}). We also exclude \textsc{Transfer Outside Europe.Adequacy Decision.\{Territory} \textit{and} \textsc{Sector}\} because we have no examples in our experimental material (both for training or testing). To summarize the different metrics, we report the \textit{micro average} across the different metadata types, by computing the metrics on all TPs, FPs, FNs, TNs found across all metadata types. 

Accuracy evaluates how the metadata identification approach performs in correctly predicting 
\textcolor{black}{the manifestations of metadata  
in the privacy policies. }
Apart from the excluded metadata types \textcolor{black}{(explained above)}, the table shows that the presence or absence of all metadata types are identified with an accuracy greater than 80\%. The relatively low accuracy in the case of \textsc{PD~Origin.Direct} and \textsc{PD Origin.Indirect.Third Party} is due to 
\textcolor{black}{eight manifestations which our approach identifies, but the sentences representing these manifestations predicted by the approach were not the same as the ones in the ground truth. 
}
We thus counted these \textcolor{black}{manifestations} 
as both FPs and FNs. 
As for \textsc{PD Provision Obliged}, the approach produces 12 errors and achieves a relatively low accuracy: $\approx$75.5\%. We note that this metadata is usually expressed in a conditional statement, e.g., if the individual fails to provide the personal data as needed, there will be consequences. Conditional sentences in English take multiple forms. Thus, semantic analysis would be required to improve the accuracy of identifying this metadata. 

Precision reflects how many actual \textcolor{black}{manifestations} 
of the metadata are correctly identified by the approach out of the total number of identified \textcolor{black}{manifestations.} 
\textcolor{black}{ Our approach achieves a precision greater than 80\% for 51 out of the 54 
metadata types.} 
In the case of metadata type \textsc{Legal Basis.Contract.To Enter Contract}, at level L3, the reason for achieving low precision is the reliance on keyword search. Keywords can easily introduce false positives \textcolor{black}{as we elaborate later in our analysis under RQ6.} 
The same reasons mentioned above for the low accuracy of \textsc{PD Provision Obliged} and \textsc{PD Origin.Indirect.Third Party} can also explain the low precision of these two metadata types. In total, our approach introduces 119 false positives out of 1449 identified \textcolor{black}{manifestations.} 

Recall assesses how many actual \textcolor{black}{manifestations 
of metadata types 
in the privacy policies are also correctly identified.} The table shows that we achieve a high recall for all metadata types, except for \textsc{Controller Representative.Contact} and \textsc{PD Provision Obliged}. Note that, in \textit{E}, we only have three \textcolor{black}{manifestations} 
of  \textsc{Controller Representative.Contact}, and the low recall (66.7\%) is due to missing only one \textcolor{black}{manifestation.} 
In total, our approach missed 68 \textcolor{black}{manifestations} 
from a total of 1448 actual \textcolor{black}{manifestations of } metadata type 
in \textit{E}. 

\textbf{The answer to RQ4} is that our metadata identification approach achieves an average accuracy, precision, recall and F2-measure of 93.4\%, 92.1\%, 95.3\% and 94.9\%, respectively. 

\begin{table*}[!t]
\centering
\caption{Results of Metadata Identification.} 
\begin{tabular}{c p{0.3\linewidth} c c c c | c c c c} 
&&\multicolumn{4}{c}{AI-based solution (\textbf{RQ4})} & \multicolumn{4}{c}{KW-based solution (\textbf{RQ6})}  \\
\midrule
\textbf{Level} & \textbf{Metadata} & \textbf{A (\%)} & \textbf{P (\%)} & \textbf{R (\%)} & \textbf{F2 (\%)} & \textbf{A (\%)} & \textbf{P (\%)} & \textbf{R (\%)} & \textbf{F2 (\%)}\\
\toprule
L1&\textbf{\textsc{Controller}}&-&-&-&-&-&-&-&-\\
L2&\textsc{Contact} & 91.8  & 94.1 & 94.1 & 94.1 
& 71.4 & 71.7 & 97.1 & 90.7 \\
L3&\textsc{Phone Number} & 95.8  & 92.9 & 92.9 & 92.9 
& 85.4 & 68.4 & 92.9 & 86.7 \\
L3&\textsc{Email} & 85.7 & 90.9 & 80.0 & 82.0 
& 62.5 & 58.1 & 100 & 87.4 \\
L3&\textsc{Legal Address} & 86.0 & 88.9 & 85.7 & 86.3
& 49.1 & 51.1 & 82.1 & 73.2 \\
\midrule
L1&\textbf{\textsc{Controller Representative}}&-&-&-&-&-&-&-&-\\
L2&\textsc{Contact} & 97.9  & 100 & 66.7 & 71.4 
& 08.2 & 04.3 & 66.7 & 17.2 \\
L3&\textsc{Legal Address} & 97.9  & 100 & 66.7 & 71.4 
& 10.2 & 04.4 & 66.7 & 17.5 \\
\midrule
L1&\textbf{\textsc{Data Subject Right}} & 97.9 & 97.8 & 100 & 99.5 
& 91.7 & 91.7 & 100 & 98.2 \\
L2&\textsc{Access} & 88.0 & 90.9 & 95.2 & 94.3 
& 84.0 & 87.0 & 95.2 & 93.5 \\
L2&\textsc{Rectification} & 100 & 100 & 100 & 100 
& 70.4 & 81.0 & 81.0 & 81.0 \\
L2&\textsc{Restriction} & 95.8 & 94.3 & 100 & 98.8 
& 93.8 & 91.7 & 100 & 98.2 \\
L2&\textsc{Complaint} & 100 & 100 & 100 & 100 
& 87.5 & 85.4 & 100 & 96.7 \\
L3&\textsc{SA} & 100 & 100 & 100 & 100 
& 60.4 & 67.6 & 73.5 & 72.3 \\
L2&\textsc{Erasure} & 100 & 100 & 100 & 100 
& 100 & 100 & 100 & 100 \\
L2&\textsc{Object} & 97.9 & 97.3 & 100 & 99.4 
& 97.9 & 97.3 & 100 & 99.4 \\
L2&\textsc{Portability} & 100 & 100 & 100 & 100 
& 95.9 & 96.9 & 96.9 & 96.9 \\
L2&\textsc{Withdraw Consent} & 95.8 & 100 & 93.9 & 95.1 
& 95.8 & 94.3 & 100 & 98.8 \\
\midrule
L1&\textbf{\textsc{Legal Basis}} & 97.9 & 97.9 & 100 & 99.6 
& 95.8 & 95.8 & 100 & 99.1 \\
L2&\textsc{Contract} & 85.4 & 82.9 & 100 & 96.0 
& 70.8 & 70.8 & 100 & 92.4 \\
L3&\textsc{Contractual} & 86.0 & 88.6 & 91.2 & 90.6 
& 83.7 & 90.6 & 85.3 & 86.3 \\
L3&\textsc{To Enter Contract} & 83.3 & 70.8 & 94.4 & 88.5 
& 79.2 & 64.3 & 100 & 90.0 \\
L3&\textsc{Statutory} & 97.9 & 100 & 85.7 & 88.2 
& 83.3 & 45.5 & 71.4 & 64.1\\
L2&\textsc{Public Function} & 83.7 & 81.8 & 81.8 & 81.8 
& 45.8 & 45.8 & 100 & 80.9 \\
L2&\textsc{Legitimate Interest} & 84.0 & 92.9 & 88.6 & 89.4
& 91.7 & 91.7 & 100 & 98.2 \\
L2&\textsc{Vital Interest} & 100 & 100 & 100 & 100 
& 14.6 & 14.6 & 100 & 46.1 \\
L2&\textsc{Consent} & 93.8 & 95.0 & 97.4 & 96.9 
& 81.3 & 81.3 & 100 & 95.6 \\
L2&\textsc{Legal Obligation} & 93.9 & 95.7 & 97.8 & 97.3 
& 93.8 & 93.8 & 100 & 98.7 \\
\midrule
L1&\textbf{\textsc{Transfer Outside Europe}} & 97.9 & 96.8 & 100 & 99.3
& 73.5 & 70.7 & 96.7 & 90.1 \\
L2&\textsc{Adequacy Decision} & 97.9 & 80.0 & 100 & 95.2 
& 66.7 & 20.0 & 100 & 55.6 \\
L3&\textsc{Country} & 100 & 100 & 100 & 100 
& 79.2 & 10.0 & 50.0 & 27.8 \\
L2&\textsc{Safeguards} & 97.9 & 95.8 & 100 & 99.1 
& 97.9 & 95.8 & 100 & 99.1 \\
L3&\textsc{Binding Corporate Rules} & 100 & 100 & 100 & 100 
& 100 & 100 & 100 & 100 \\
L3&\textsc{EU Model Clauses} & 97.9 & 95.2 & 100 & 99.0 
& 97.9 & 95.2 & 100 & 99.0 \\
L2&\textsc{Specific Derogation} & 97.9 & 87.5 & 100 & 97.2 
& 60.4 & 20.0 & 57.1 & 41.7 \\
L3&\textsc{Unambiguous Consent} & 97.9 & 85.7 & 100 & 96.8 
& 58.3 & 15.0 & 50.0 & 34.1 \\
\midrule
L1&\textbf{\textsc{PD Origin}} & 93.8 & 93.8 & 100 & 98.7 
& 93.8 & 93.8 & 100 & 98.7 \\
L2&\textsc{Direct} & 76.5 & 80.4 & 92.5 & 89.8 
& 83.3 & 83.3 & 100 & 96.2 \\
L2&\textsc{Indirect} & 93.9 & 95.7 & 97.8 & 97.3 
& 93.8 & 93.8 & 100 & 98.7  \\
L3&\textsc{Third Party} & 71.7 & 76.5 & 78.8 & 78.3 
& 45.8 & 51.6 & 48.5 & 49.1 \\
L3&\textsc{Publicly} & 89.6 & 82.8 & 100 & 96.0
& 83.3 & 75.0 & 100 & 93.8 \\
L3&\textsc{Cookie} & 95.8 & 94.1 & 100 & 98.8 
& 89.9 & 90.9 & 93.8 & 93.2 \\
\midrule
L1&\textbf{\textsc{PD Category}} & 93.9 & 95.7 & 97.8 & 97.4 
& 62.0 & 93.5 & 63.0 & 67.4 \\
L2&\textsc{Special} & 95.8 & 93.5 & 100 & 98.6 
& 95.8 & 93.5 & 100 & 98.6 \\
L2&\textsc{Type} & 81.6 & 50.0 & 77.8 & 70.0 
& 81.3 & 0.0 & 0.0 & 0.0 \\
\midrule
L1&\textbf{\textsc{Recipients}} & 93.8 & 93.3 & 100 & 96.6 
& 29.0 & 41.9 & 42.9 & 42.7 \\
\midrule
L1&\textbf{\textsc{PD Time Stored}} & 91.7 & 94.7 & 94.7 & 94.7 
& 86.0 & 87.8 & 94.7 & 93.3 \\
\midrule
L1&\textbf{\textsc{PD Provision Obliged}} & 75.5 & 76.9 & 76.9 & 76.9 
& 54.2 & 54.2 & 100 & 85.5 \\
\midrule
L1&\textbf{\textsc{Processing Purposes}} & 84.3 & 91.3 & 91.3 & 91.3 
& 40.0 & 58.1 & 54.3 & 55.1 \\
\midrule
L1&\textbf{\textsc{PD Security}} & 82.7 & 88.4 & 90.5 & 90.0 
& 83.7 & 85.4 & 97.6 & 94.9 \\
\midrule
L1&\textbf{\textsc{Auto Decision Making}} & 93.8 & 95.0 & 90.5 & 91.3 
& 64.6 & 55.3 & 100 & 86.1 \\
\midrule
L1&\textbf{\textsc{Children}} & 95.8 & 80.0 & 80.0 & 80.0 
& 77.1 & 31.3 & 100 & 69.4 
\\
\midrule
L1&\textbf{\textsc{DPO}} & - & - & - & - & - & - & - & - \\
L2&\textsc{Contact} & 97.9 & 95.5 & 100 & 99.1 
& 47.9 & 45.7 & 100 & 80.8 \\
L3&\textsc{Phone Number} & 100  & 100 & 100.0 & 100.0 
& 62.5  & 5.6 & 50.0 & 19.2 \\
L3&\textsc{Email} & 97.9 & 95.0 & 100 & 99.0 
& 50.0 & 44.2 & 100 & 79.8 \\
L3&\textsc{Legal Address} & 91.8  & 81.3 & 92.9 & 90.3 
& 20.4 & 17.8 & 57.1 & 39.6 \\
\midrule
&\textbf{Summary}& 93.4 & 92.1 & 95.3 & 94.6 
& 70.7 & 65.2 & 90.1 & 83.7 \\ 
\bottomrule
\end{tabular}
\label{tab:resultsOfMetadaIdentification}
\end{table*}

\vspace{.5em}

\sectopic{RQ5. How accurate is our approach in checking the completeness of privacy policies?}

Table~\ref{tab:resultsOfCompletenessChecking}, on the left-hand side,  shows the results of EXPII. We evaluate in RQ5 how well our completeness criteria (see Sec.~\ref{sec:criteria}) can detect incompleteness issues, given the metadata identified by the approach and evaluated in RQ4. An incompleteness issue can be either a  violation or a  warning \textcolor{black}{(as defined in Sec.~\ref{sec:criteria})}. 
The table reports the number of TPs, FPs, FNs and TNs \textcolor{black}{(redefined for EXPII in Sec.~\ref{subsec:evalProcedure})} in addition to the evaluation metrics, namely accuracy (A), precision (P), recall (R) and F2-measure (F2). 

We note that seven criteria lead to warnings, namely C6, C11 -- C14, C16 and C18. The remaining criteria lead to violations. We also note that C1, C2, C5, C20, and C21 are concerned with the unconditional presence of mandatory metadata types, whereas the criteria C3, C4, C10 -- C19, C22 and C23 need to be checked only in specific situations based on the answers provided on the questionnaire (explained in Sec.~\ref{sec:criteria}). For the latter set of criteria, we assume in our evaluation that they  \textit{always} need to be checked. The criteria C6 -- C9, C11 -- C14, C16, C18 and C22 are concerned with 
\textcolor{black}{metadata types that need to be present only if some other metadata types are also present in the same privacy policy. }

\sectopic{Violations.} Out of 285 genuine violations in the test privacy policies, our completeness criteria correctly detect 261, while introducing 16 false positives. This results in a precision of 94.2\% and a recall of 91.6\%. 


Table~\ref{tab:resultsOfCompletenessChecking} shows that the approach introduces 40 errors (16 FPs and 24 FNs) that led to violations. We analyzed the reason for having these errors. Out of the 40 errors, 26 are originated from false positives in the metadata identification results. \textcolor{black}{The remaining 14 errors are due to missed metadata types (FNs) across seven criteria. Specifically, one or two missed metadata types yielded errors in \textbf{C02}, \textbf{C04}, \textbf{C08}, \textbf{C09} and \textbf{C21}, whereas five missed metadata types yielded errors in \textbf{C22}. }
The low precision and recall values for the completeness checking of \textbf{C22} are in part due to the accuracy of identifying \textsc{PD Provision Obliged}. 
As we explained in RQ4, this metadata type requires further analysis to capture the variations of how it is expressed. 

\begin{table*}[!t]
\centering
\caption{Results of Completeness Checking. } 
\begin{tabular}{p{0.05\linewidth} c c c c | c c c c 
| c c c c 
| c c c c
} 
&\multicolumn{7}{c}{AI-based approach (\textbf{RQ5})} & \multicolumn{7}{c}{KW-based baseline (\textbf{RQ6})}\\
\midrule
\textbf{Criteria} & \textbf{TPs} & \textbf{FPs} & \textbf{FNs} & \textbf{TNs} & \textbf{A (\%)} & \textbf{P (\%)} & \textbf{R (\%)} & \textbf{F2 (\%)} 
& \textbf{TPs} & \textbf{FPs} & \textbf{FNs} & \textbf{TNs} & \textbf{A (\%)} & \textbf{P (\%)} & \textbf{R (\%)} & \textbf{F2 (\%)} 
\\
\toprule
\colorbox{red!25}{C01} & 2 & 0 & 0 & 46 & 100 & 100 & 100 & 100 
& 2 & 0 & 0 & 46 & 100 & 100 & 100 & 100 
\\
\colorbox{red!25}{C02} & 13 & 1 & 1 & 33 & 95.8 & 92.9 & 92.9 & 92.9 
& 2 & 0 & 12 & 34 & 75.0 & 100 & 14.3 & 17.2 
\\
\colorbox{red!25}{C03} & 45 & 0 & 0 & 3 & 100 & 100 & 100 & 100 
& 42 & 1 & 3 & 2 & 91.7 & 97.7 & 93.3 & 94.2 
\\
\colorbox{red!25}{C04} & 45 & 1 & 0 & 2 & 97.9 & 97.8 & 100 & 99.6 
& 2 & 0 & 43 & 3 & 10.4 & 100 & 04.4 & 05.5 
\\
\colorbox{red!25}{C05} & 36 & 0 & 4 & 152 & 97.9 & 100 & 90.0 & 91.8 
& 24 & 2 & 15 & 152 & 91.1 & 92.3 & 61.5 & 65.9 
\\ 
\colorbox{red!25}{C07} & 7 & 4 & 0 & 37 & 91.7 & 63.6 & 100 & 89.7 
& 7 & 9 & 0 & 32 & 81.3 & 43.8 & 100 & 79.5 
\\
\colorbox{red!25}{C08} & 7 & 1 & 3 & 37 & 91.7 & 87.5 & 70.0 & 72.9 
& 9 & 2 & 1 & 36 & 93.8 & 81.8 & 90.0 & 88.2 
\\
\colorbox{red!25}{C09} & 31 & 1 & 1 & 159 & 99.0 & 96.0 & 96.9 & 96.9 
& 30 & 19 & 2 & 141 & 89.1 & 61.2 & 93.8 & 84.7 
\\
\colorbox{red!25}{C10} & 17 & 0 & 1 & 30 & 97.9 & 100 & 94.4 & 95.5 
& 7 & 0 & 11 & 30 & 77.1 & 100 & 38.9 & 44.3 
\\
\colorbox{red!25}{C15} & 2 & 0 & 1 & 45 & 97.9 & 100 & 66.7 & 71.4 
& 0 & 0 & 3 & 45 & 93.8 & n/a & 0.0 & n/a 
\\
\colorbox{red!25}{C17} & 1 & 0 & 1 & 46 & 97.9 & 100 & 50.0 & 55.6 
& 2 & 15 & 0 & 31 & 68.8 & 11.8 & 100 & 40.0 
\\
\colorbox{red!25}{C19} & 3 & 0 & 3 & 42 & 93.8 & 100 & 50.0 & 55.6 
& 2 & 3 & 4 & 39 & 85.4 & 40.0 & 33.3 & 34.5 
\\
\colorbox{red!25}{C20} & 8 & 2 & 2 & 36 & 91.7 & 80.0 & 80.0 & 80.0 
& 7 & 0 & 3 & 38 & 93.8 & 100 & 70.0 & 74.5 
\\
\colorbox{red!25}{C21} & 1 & 1 & 1 & 45 & 95.8 & 50.0 & 50.0 & 50.0 
& 1 & 4 & 1 & 42 & 89.6 & 20.0 & 50.0 & 38.5 
\\
\colorbox{red!25}{C22} & 17 & 5 & 5 & 21 & 79.2 & 77.3 & 77.3 & 77.3 
& 0 & 0 & 22 & 26 & 54.2 & n/a & 0.0 & n/a 
\\
\colorbox{red!25}{C23} & 26 & 0 & 1 & 21 & 97.9 & 100 & 96.3 & 97.0 
& 2 & 0 & 25 & 21 & 47.9 & 100 & 07.4 & 9.1 
\\
\colorbox{orange!25}{C06} & 1 & 0 & 0 & 47 & 100 & 100 & 100 & 100 
& 0 & 4 & 1 & 43 & 89.6 & 0.0 & 0.0 & n/a 
\\
\colorbox{orange!25}{C11} & 6 & 0 & 0 & 42 & 100 & 100 & 100 & 100 
& 1 & 0 & 5 & 42 & 89.6 & 100 & 16.7 & 20.0 
\\
\colorbox{orange!25}{C12} & 2 & 1 & 0 & 45 & 97.9 & 66.7 & 100 & 90.9 
& 2 & 4 & 0 & 42 & 91.7 & 33.3 & 100 & 71.4 
\\
\colorbox{orange!25}{C13} & 3 & 0 & 0 & 45 & 100 & 100 & 100 & 100 
& 3 & 0 & 0 & 45 & 100 & 100 & 100 & 100 
\\
\colorbox{orange!25}{C14} & 1 & 0 & 0 & 47 & 100 & 100 & 100 & 100 
& 0 & 0 & 1 & 47 & 97.9 & n/a & 0.0 & n/a 
\\
\colorbox{orange!25}{C16} & 6 & 1 & 3 & 38 & 91.7 & 85.7 & 66.7 & 69.8 
& 5 & 4 & 4 & 35 & 83.3 & 55.6 & 55.6 & 55.6 
\\
\colorbox{orange!25}{C18} & 20 & 5 & 7 & 16 & 75.0 & 80.0 & 74.1 & 75.2 
& 24 & 15 & 3 & 6 & 62.5 & 61.5 & 88.9 & 81.6 
\\
\midrule
\textbf{Summary} & 300 & 23 & 34 & 1035 & 95.9 & 92.9 & 89.8 & 90.4 
& 174 & 82 & 159 & 977 & 82.7 & 68.0 & 52.3 & 54.8 
\\
\bottomrule
\end{tabular}
\label{tab:resultsOfCompletenessChecking}
\end{table*}

\sectopic{Warnings.} Out of 49 genuine warnings in the test privacy policies, our completeness criteria correctly detect 39, while introducing seven false positives. This results in a precision of 84.4\% and a recall of 79.6\%. A total of 17  errors resulted in warnings, including seven FPs and 10 FNs. All of these errors are due to FPs from metadata identification, except one that is due to a missed metadata in one privacy policy. 

\textcolor{black}{Our completeness checking approach generates a report, as an output, that is shared with the analyst. The report includes not only the final decision regarding whether a privacy policy is or is not complete according to GDPR, but also a structured summary of the identified metadata types.  \textcolor{black}{Specifically, the report lists under each criterion the set of sentences describing the identified 
manifestations of the metadata types or ``not found'' in case no manifestations 
are found by our approach.} The analyst can then review this summary instead of analyzing a privacy policy in its entirety. The criteria that erroneously result in violations or raise warnings due to false positives in the metadata identification (33 in total) can be easily filtered out by the analyst. 
If the analyst reviews the incompleteness issues only instead of the summary, our approach still fares well in identifying about 90\% of the actual violations and warnings. In practice, the accuracy of our approach is sufficient to be used by diverse users, including software engineers who might lack legal expertise or legal experts who need assistance to optimize \hbox{their time and effort.  } }


Based on our assumption that all criteria need to be satisfied in order for a privacy policy to be complete according to GDPR, all of the 48 policies in the test set are incomplete. 
\textcolor{black}{Our approach is able to correctly
identify all the privacy policies that have some incompleteness
issue.} 

\textbf{The answer to RQ5} is that our completeness checking approach can detect incompleteness issues in privacy policies with an average accuracy, precision, recall and F2-measure of 95.9\%, 92.9\%, 89.8\% and 90.4\%, respectively.

\vspace{.5em}
\sectopic{\textcolor{black}{RQ6. Is our approach worthwhile compared to a simpler solution?}} 

Tables~\ref{tab:resultsOfMetadaIdentification}~and~\ref{tab:resultsOfCompletenessChecking} show the results of EXPIII including, on the left-hand side, the results of our AI-based approach as discussed in RQ4 and RQ5, and on the right-hand side the results of the KW-based approach that we introduced in Sec.~\ref{subsec:evalProcedure}. 

\sectopic{Metadata identification.} The table suggests that there are two disadvantages of using the KW-based approach. First, not all of the metadata types can be accurately identified using keywords, e.g., \textsc{Recipients}. Recall our discussion in Sec.~\ref{sec:introduction} about this metadata type that can include a list of organizations. \textcolor{black}{A finite list of predefined keywords cannot possibly cover all organization names that might appear in \textsc{Recipients} or capture the diverse \textsc{Processing Purposes} of personal data. }
\textcolor{black}{Our ground truth contains a total of 42 actual manifestations 
of \textsc{Recipients} and 46 of \textsc{Processing Purposes}. We note that 21 manifestations 
of \textsc{Recipients} and 17 manifestations 
of \textsc{Processing Purposes} predicted by KW-based are counted as both FPs and FNs, because none of the identified sentences associated with these manifestations 
are matching the ones in the ground truth. In contrast, our AI-based approach finds only three manifestations 
of \textsc{Processing Purposes} with irrelevant sentences. This shows that our approach is more reliable in finding the correct sentences related to the identified manifestations 
(35 less such errors). }
  

The second disadvantage of KW-based is that, though it achieves a relatively good recall, this comes at the cost of precision.  For example, the recall for identifying the metadata type \textsc{Transfer Outside Europe.Adequacy Decision} using keywords is 100\% but the precision is only 20\%. Despite such high recall, our AI-based approach achieves an overall better F2-measure, namely +11\%.
\textcolor{black}{To summarize, our AI-based approach misses in total 68 manifestations of metadata types 
(FNs)} and introduces 119 FPs, whereas KW-based misses 144 FNs and produces 697 FPs (76 more FNs and 578 more FPs than our approach). As a result, we achieve a gain of $\approx$5\% in recall and $\approx$27\% in precision. 

\sectopic{Completeness checking.} The difference in performance becomes even clearer in the completeness checking task, which depends largely on the accuracy of metadata identification. The KW-based approach has a respective precision and recall of 71.6\% and 48.9\% for detecting violations, and 55.7\% and 69.4\% for detecting warnings. 
\textcolor{black}{In comparison with the total of 57 errors produced by our approach for violations and warnings, KW-based produces 241 errors (i.e., 184 more errors). Of these 241 errors, 45 are due to missed \textcolor{black}{manifestations of metadata types 
(FNs),}  15 are caused by \textsc{PD Category} (which is hard to capture via keywords), and the remaining 196 are originating from false positives in metadata identification. Filtering so many cases out is, compared to the 33 FPs introduced by our approach, much more time-consuming for the analyst.
Our approach is therefore advantageous over the KW-based solution in terms of both precision and recall.} 
Specifically, \textcolor{black}{using a combination of NLP and ML} 
leads to a significant improvement of $\approx$23\% in precision and $\approx$43\% in recall for detecting violations. The overall improvement, considering both warnings and violations, is $\approx$25\% higher precision, $\approx$38\% higher recall, and $\approx$36\% higher F2-measure. 

\textbf{The answer to RQ6} is that our AI-based approach presents a significant improvement over merely using keywords, both in metadata identification and in completeness checking. In metadata identification, our approach outperforms the KW-based solution by an average of 22.7\% in accuracy, 26.9\% in precision, 5.2\% in recall and 11\% in F2-measure. This leads to a significant follow-on gain in completeness checking, where our approach outperforms the baseline by 24.5\% in precision and 38\% in recall. 

\section{Threats to Validity} \label{sec:threats}

Below, we discuss threats to the validity of our empirical results and what we did to mitigate these threats.

\textbf{Internal Validity.} Bias was an important concern in relation to internal validity. To mitigate bias, we curated most ($\approx$90\%) of the manual annotations through third-parties (non-authors). Another potential threat to internal validity is that the authors interpreted the text of GDPR provisions in order to create the privacy policy conceptual model presented in Fig.~\ref{fig:metadata}. 
\textcolor{black}{To minimize the threat posed by a subjective interpretation, this phase was done in close collaboration with three independent legal experts from Linklaters, who have expertise in European and international laws with a focus on the data protection and financial domains.  
While we cannot entirely rule out subjectivity, we provide our interpretation in a precise and explicit form. In addition, our model is publicly available and thus open to scrutiny.}
\textcolor{black}{Another threat to internal validity is our reliance on a static set of keywords. Changing this set might have an impact on the results of our automated solution. However, we believe that our set of keywords is reasonably adequate and complete since we manually created the keywords during our qualitative study in close collaboration with legal experts. 
}
\textbf{External Validity.} The qualitative study through which we built our conceptual model of privacy policies is domain-agnostic: the study was rooted in GDPR and further enhanced by feedback from legal experts who had familiarity with data protection in a variety of domains. This provides a fair degree of confidence about our conceptual model being generalizable.  As for our evaluation of automation accuracy of our completeness checking approach (see Sec.~\ref{subsec:RQs}), and more specifically, whether the accuracy levels observed would generalize beyond the fund domain, we note that certain metadata types were rare in privacy policies from the fund domain. Furthermore, we have not yet conducted a multi-domain evaluation of our metadata identification and completeness checking approaches. For these reasons, it would be premature to make claims about how our accuracy results would carry over to other domains. That said, we believe that the core components of our automation approach, notably, our hybridized use of word embeddings, ML-based classification, similarity analysis and keyword search, provides a versatile basis for the future development of a more broadly applicable solution to check the completeness of privacy policies.


\section{Related Work} \label{sec:related}

\textcolor{black}{Our proposed approach for checking the completeness of privacy policies spans three different tasks. The first task involves the \textcolor{black}{elicitation} of privacy-related requirements for GDPR compliance. The second task covers the completeness checking of privacy policies (with a GDPR focus). The last task is concerned with checking the data handling practices and privacy compliance of software against their associated privacy policies. This last task enables an implicit compliance checking of the software against the privacy-requirements stated in the provisions (GDPR, in our case). 
Our work concentrates on  providing automation for the first two tasks, noting that the results from these two tasks also serve as an input to the third task, which we do not directly address in this article. Below, we position our work against the related work
on (i) identifying privacy-policy requirements, (ii) completeness checking of privacy policies, and (iii) completeness/compliance checking of software against data protection regulations.}

\subsection{\textcolor{black}{Elicitation of Privacy-related Requirements}} 
Vanezi et al. \cite{VaneziKKPP20} propose a graphical modeling language for GDPR privacy policies and a methodology for transforming such graphically-defined privacy policies into formal definitions. This work focuses on one (namely, \textsc{Processing Purposes}) out of the 56 metadata types we consider in our work. Caramujo et al.~\cite{Caramujo2019} target privacy policies for the web and mobile applications, and propose a domain-specific language along with model transformations for specifying privacy-policy models. Similarly, Pullonen et al.~\cite{Pullonen2019} present a multi-level model to be used as an extension of the Business Process Model and Notation to enable the visualization, analysis, and communication of the privacy-policy characteristics of business processes. Finally, Kumar and Shyamasundar ~\cite{Kumar14} explore the suitability of information-flow controls as a tool for specifying and enforcing privacy-policy requirements.
These existing works address a rather small subset of the privacy-policy metadata types considered in this article. In addition, excluding~\cite{VaneziKKPP20}, all of the above-mentioned papers focus on providing guidelines that are not strictly based on GDPR. In contrast, we systematically identify the requirements that, according to GDPR, must be met by privacy policies for completeness.

\subsection{Completeness Checking of Privacy Policies} 

{\color{black} S\'{a}nchez et al.~\cite{Sanchez21} check the compliance of privacy policies with respect to  six data protection goals as stated by GDPR, including lawfulness, purpose limitation, data minimisation, accuracy, storage limitation, and integrity and confidentiality. The authors use four privacy policies to train binary classifiers for deciding whether a privacy policy is compliant with respect to each of the six goals. These goals cover only 15 out of the 56 metadata types we handle.

Nejad et al.~\cite{Nejad20} present three different models for classifying the paragraphs of privacy policies into pre-defined categories using supervised machine learning. To train their models, the authors use a dataset containing 115 privacy policies from various US companies. The authors consider 12 high-level categories for their classification. All these categories are included in our set of metadata types. 
}

Tesfay et al.~\cite{Tesfay2018} propose a ML-based method for classifying the content of privacy policies across 10 categories using predefined keywords. Those categories are all covered by our metadata types, except for one category, Policy Change, which is orthogonal to our purposes.

Bhatia et al.~\cite{Bhatia2016} develop a semi-automated framework for extracting privacy goals from privacy policies through crowdsourcing and NLP.  Similar crowdsourcing initiatives have been proposed by others as well, e.g., by Liu et al.~\cite{Liu2014} and Wilson et al.~\cite{Wilson2016}, where privacy policies are manually annotated in order to match their text segments against privacy issues of interest. Guerriero et al.~\cite{Guerriero18} propose a framework for specifying, enforcing and checking privacy policies in data-intensive applications.
Bhatia et al. ~\cite{Bhatia19} present a semantic frame-based representation for privacy statements that can be used to identify incompleteness in four categories of data action: collection, retention, usage, and transfer. 
Lippi et al.~\cite{lippi:19} present 33 metadata types for GDPR privacy policies and provide automatic support for vagueness detection based on manually crafted rules and  
ML classifiers built using the exact terminology of the policies as learning features. 

In summary, in comparison to the above-cited works, we have a different analytical focus, namely completeness checking. In terms of the metadata types, our proposed 56 types cover all the ones identified by others, except -- as noted before -- one metatadata type, Policy Change~\cite{Tesfay2018}, which is orthogonal to completeness checking. 
Furthermore, the existing approaches outlined above rely to a large extent on the exact phrasing of the policies to be able to extract and classify information. They do not present a thorough conceptualization of the content expected in privacy policies. The scope of application of these approaches is thus limited and, where automation is provided, the accuracy is not high enough for industrial use. 
In this article, we addressed the above limitations by considering a wider set of metadata types and using a combination of advanced NLP and ML for automated support.

\textcolor{black}{
\subsection{Compliance Checking of Software}
Fan et al.~\cite{Fan:20} check for the compliance of mobile health applications against GDPR. To do so, the authors propose an automated system for detecting three types of violations: incompleteness of the app privacy policy, inconsistency of data collection, and non-secure data transmission. For incompleteness checking of privacy policies, the authors define six categories \textcolor{black}{of privacy-related information} that need to be present in a privacy policy. They apply ML-based binary classifiers on bag-of-words representations of the sentences in a given policy to predict whether any of the categories is present in the policy. Specifically, they apply random forest (RF), decision trees (DT) and na\"ive Bayes (NB). Based on 10-fold cross-validation over 100 privacy policies (1,284 labelled sentences), RF performed the best in four of the six categories, and DT performed the best in the other two. 
The best reported precision and recall are on average 92.5\% and 93.3\%, respectively. 
In contrast with our work, the authors consider only six out of the 56 metadata types that we present in this article. Moreover, and from an ML standpoint, our solution architecture is different: we use embeddings to create the representations of the text in a given policy and apply an ensemble classification approach. }

\textcolor{black}{The COVID-19 pandemic  has heightened privacy concerns for individuals, as seen, for example, in the analysis of app reviews for COVID-19 contact-tracing apps~\cite{Bano:21}. }
\textcolor{black}{Hatamian et al.~\cite{Hatamian:21} analyze the privacy and security aspects of COVID-19 contact tracing apps.
In their analysis, they consider 12 metadata types derived from different GDPR articles, including children protection, data retention and others.
The authors collect the data access intentions from the permissions an Android app is given, e.g., the access to data such as call logs and contact lists. Through manual incompleteness checking of the privacy policies of 28 COVID-19 contact tracing apps, the authors assess  the extent to which the policies cover the 12 GDPR principles. Subsequently, the authors check whether the apps fulfill the provisions in their respective privacy policies. 
Nine of the privacy principles addressed by 
Hatamian et al. are pertinent to privacy-policy completeness checking and are thus tackled in our work. However, we provide a more elaborate treatment of these principles (metadata types). Moreover, we devise an AI-based solution to automatically identify these metadata types in the privacy policies and thereby \hbox{analyze incompleteness.}} 

\textcolor{black}{Kununka et al.~\cite{Kununka:17} assess the compliance of Android and iOS apps with their privacy policies. The basis for selecting which apps to analyze is the number of third-party domains that the apps transfer sensitive data to. In total, 30 apps are selected. The authors first analyze the categories of personal data transferred to a third-party. They then manually identify metadata types in the privacy policies of these apps, focusing only on the collection, use and transfer of personal data. Finally, the authors check for the compliance of the data practices in the apps against what is stated in the policies. 
Compared to our automated approach, their metadata identification from both the apps and their privacy policies, as well as the compliance checking, are done entirely manually. Further, they consider only two metadata types (i,e., \textsc{PD Category} and its specialization \textsc{Special}) from what we present in this article.}

{\color{black}
\section{Replicating our Methodology}\label{sec:reusability}
Our proposed completeness checking process is not limited to privacy policies and GDPR, and can be instantiated for checking the completeness of any given document type (\textit{D}) according to any given regulation (\textit{R}). In our context, \textit{D} represents a privacy policy, and \textit{R} is GDPR. Reusing our approach can be done by replicating the same methodology as described in this paper. Specifically, one must first conduct a qualitative study over those of  \textit{R}'s provisions that are relevant to checking the completeness of \textit{D}. Such a qualitative study should aim at building a conceptual model and a set of completeness checking criteria.  Subsequently, one must develop automation for completeness checking. When supervised machine learning is used for automation, one will need to (manually) create a labeled dataset covering a relatively large number of documents of type \textit{D}. This will be followed by the development of classification methods, potentially alongside prediction rules and post-processing steps.  

The effort required to replicate our methodology for other regulations and document types depends on several factors, including the number and complexity of the provisions that need to be considered in the qualitative study, the background and expertise in conceptual modeling and AI, the size of the evaluation data and the complexity of the classification algorithms used in the work. 
Based on our experience, we anticipate that 30-40\% of the effort would go towards building a conceptual model and completeness criteria and the remaining 60-70\% would go towards developing an automated solution.
}

\section{Conclusion} \label{sec:conclusion}

In this paper, we proposed an AI-enabled approach for completeness checking of privacy policies according to the General Data Protection Regulation (GDPR). We first developed a conceptual model aimed at providing a thorough characterization of the content of privacy policies. Based on this conceptual model, we devised criteria describing how a privacy policy should be checked for completeness against GDPR. Second, using Natural Language Processing and Machine Learning, we developed automated support for classifying the content of privacy policies and thus identifying the metadata types necessary for checking privacy-policy completeness.


We curated a considerable number of annotated privacy policies (234 policies in total), with the majority of the annotation work performed by third-parties. We evaluated our approach 
on a test set of 48 privacy policies. 
Our metadata identification approach achieved an average precision of 92\% with an average recall of 95\% for identifying the \textcolor{black}{manifestations} 
of all metadata types across the test privacy policies. We ran the completeness criteria over the identified metadata. Our completeness checking approach was able to detect 300 out of 334 incompleteness issues in the real-world privacy policies we used for validation. The approach also generated 23 false positives. Our completeness checking approach thus had a precision of 93\% and a recall of 90\% over our test set. Compared to an intuitive automated solution based on keyword search, our AI-based approach leads to a significant improvement in precision and recall of 27\% and 5\% for metadata identification and of 24.5\% and 38\% in completeness checking, respectively.  



In the future, we plan to enhance our completeness criteria so that they consider not only the presence/absence of metadata but also the meaning of the sentences containing the metadata.
Another important direction for future work is to go beyond our current case-study domain (funds) in order to assess the generalizability of our approach. 

\section*{Acknowledgments}

This paper was supported by Linklaters, Luxembourg's National Research Fund (FNR) under grant BRIDGES/19/IS/13759068/ARTAGO, and NSERC of Canada under the Discovery, Discovery Accelerator and CRC programs.

\bibliographystyle{IEEEtran}
\balance
\bibliography{ref}

\end{document}